\documentclass{emulateapj}	

\usepackage{amsmath}
\usepackage{url}

\begin{document}

\shorttitle{Achromatizing a liquid-crystal spectropolarimeter}
\shortauthors{Harrington, Kuhn, Sennhauser, Messersmith \& Thornton}

\title{Achromatizing a liquid-crystal spectropolarimeter: Retardance vs Stokes-based calibration of HiVIS.}
\author{D. M. Harrington$^1$, J.R. Kuhn$^1$, C. Sennhauser$^2$, E.J. Messersmith$^1$ \& R.J. Thornton$^3$}
\affil{Institute for Astronomy, University of Hawaii, Honolulu-HI-96822}
\affil{Institute for Astronomy, ETH Zurich, 8093 Zurich, Switzerland}
\email{dmh@ifa.hawaii.edu, kuhn@ifa.hawaii.edu}
\vspace{-3mm}

\begin{abstract}

	Astronomical spectropolarimeters can be subject to many sources of systematic error which limit the precision and accuracy of the instrument. We present a calibration method for observing high-resolution polarized spectra using chromatic liquid-crystal variable retarders (LCVRs). These LCVRs allow for polarimetric modulation of the incident light without any moving optics at frequencies $\geq$10Hz. We demonstrate a calibration method using pure Stokes input states that enables an achromatization of the system. This Stokes-based deprojection method reproduces input polarization even though highly chromatic instrument effects exist. This process is first demonstrated in a laboratory spectropolarimeter where we characterize the LCVRs and show example deprojections. The process is then implemented the a newly upgraded HiVIS spectropolarimeter on the 3.67m AEOS telescope. The HiVIS spectropolarimeter has also been expanded to include broad-band full-Stokes spectropolarimetry using achromatic wave-plates in addition to the tunable full-Stokes polarimetric mode using LCVRs. These two new polarimetric modes in combination with a new polarimetric calibration unit provide a much more sensitive polarimetric package with greatly reduced systematic error. 		
\end{abstract}


\section{Introduction}

	Stellar high resolution spectropolarimetry is an underutilized and often complex technique, in part because polarimetric signatures are often less than 1\% and in many cases less than 0.1\%. Nevertheless, when it can be reliably applied it is a powerful remote sensing diagnostic. In this small-signal regime, where high signal-to-noise is essential, there are important instrumental limits to the derived accuracy and precision. 
	
	There are several methodologies used for making precision measurements. The most obvious but practically most difficult method is to build a specialized instrument (and telescope) specifically optimized for polarimetry. One minimizes the number of optical elements, keeps all optical folds to near-normal incidence, minimizes scattered light and has some form of modulation or beam-swapping to remove systematic effects. 
	
	A spectropolarimeter has many functional components essential for sensitive linear or circular polarization measurements. For example, polarizing beamsplitters, fixed and variable waveplates, and specialized synchronous detectors are common sources of polarimetry-specific systematic errors. These errors include detector variations (pixel-to-pixel efficiency and field illumination effects), changes in illumination during the exposures comprising a measurement (telescope guiding error, beam wander from moving optics, atmospheric transparency variations and seeing) and any polarimetric or chromatic effects induced by the entire optical system. Some of these are easily correctable and some are not. Typically, one or two retarders are inserted in the beam to modulate the incident polarization for detection. These can be several types of achromatic retarders, Fresnel rhombs, liquid crystal variable retarders (either ferro-electric or neumatic) or piezo-elastic modulators. In some cases, the instrument, retarder or analyzer rotates to accomplish the modulation. Various choices of retardance have been used for various applications to optimize sensitivity to different types of incident polarization or for different expected signatures. Since typical polarimetric detections at these small levels are often limited by instrument systematic effects, careful designs must mitigate as many error sources as possible.
	
	Notable examples are the ESPaDOnS fiber-fed spectrograph on the 3.6m Canada France Hawaii Telescope (CFHT) and it's copy Narval on the 2m Telescope Bernard Lyot (TBL) telescope (cf. Semel et al. 1993, Donati et al. 1999, Manset \& Donati 2003). These instruments work at a resolution R$\sim$65,000 covering $\sim$3700-10500{\AA} and are in active use by many (cf. MAPP\footnote{$http://lamwws.oamp.fr/magics/mapp/FrontPage$} or MiMES\footnote{ $www.physics.queens.ca/\sim wade/mimes/$} ). The two telescopes are equatorial and the polarimetric modulation is done immediately after a collimating lens and an atmospheric dispersion corrector (ADC). The instrument is fiber-fed and has a time-variable continuum polarization at the several percent level (ESPaDOnS Instrument Website \footnote{www.cfht.hawaii.edu/Instruments/Spectroscopy/Espadons/}). A similar continuum polarization is seen in the William Wehlau Spectropolarimeter unit with a similar design to the ESPaDOnS package (Eversberg et al. 1998). The transmissive optics (lens and ADC) currently induce a time-variable cross-talk of a few percent, but for most spectropolarimeters this is quite benign. Lower spectral resolution instruments, such as HPOL and ISIS, typically have more stable polarimetric properties that have yet to be seen at higher spectral resolution (Wolff et al. 1996, ISIS Spectropolarimetry Manual Tinbergen \& Rutten 1997 \footnote{ $www.ing.iac.es/Astronomy/observing/manuals/$}). 

	Ultimately it is calibration of the telescope and instrument that achieves the highest possible system polarimetric accuracy. Even though it is difficult to keep instrumental polarization below 1\%, the effort in building a low polarization system isn't lost, as the ultimate calibrated polarization performance improves multiplicatively. Thus, sensitive techniques which determine the Mueller matrix of the telescope/polarimeter (cf. Beck et al. 2005a \& b, Kuhn et al. 1994, Joos et al. 2006, Patat \& Romaneillo 2006, Tinbergen 2007) will achieve a lower overal system noise when the raw instrumental errors are minimized. 
	
	 High-resolution astronomical spectropolarimeters have common design elements. Typically, rotating achromatic retarders are placed before a calcite-based dual-beam analyzer - a Wollaston prism in collimated space or a Savart plate at an image plane. ESPaDOnS and Narval use two half-wave and one quarter-wave Fresnel rhombs before a Wollaston prism. Two other dual-beam instruments, PEPSI on the 8.4m LBT at R$\sim$310,000 and HARPS on the ESO 3.6m telescope at R$\sim$115,000 are in various stages of construction (Strassmeier et al. 2003 \& 2008, Snik et al. 2008). 
	 
	 The HARPS polarimetric package was required to fit inside an already-existing space. This required the analyzer to be a Foster prism with a cylindrical lens on one beam and a CaF$_2$ prism for the second beam. The design includes separate quarter-wave and half-wave super-achromatic plates  for circular and linear polarization that cannot be used simultaneously.  
	 
	 The PEPSI design is similar to ESPaDOnS in that a lens collimates the telescope focus and retarders with a Wollaston perform the modulation. Another lens in combination with an atmospheric dispersion corrector form an image on fibers which feed the spectrograph. Instead of Fresnel rhombs, a super-achromatic quarter-wave plate is chosen. Linear polarization sensitivity is achieved by physically rotating the entire polarization package. 
	 
	 The High-resolution Visible and Infrared Spectrograph (HiVIS) we discuss here was an instrument initially built for spectroscopy but modified for linear spectropolarimetry (Thornton et al. 2003, Harrington et al. 2006, Harrington \& Kuhn 2008). A Savart plate and a rotating achromatic half-wave plate were installed at the entrance slit. One of the main complications of this system is the eight oblique reflections between the sky and the polarization analyzer. Since the 3.67m Advanced Electro-Optical System (AEOS) telescope is altitude-azimuth and HiVIS is in a coud\'{e} path, the many oblique reflections cause pointing-variable cross-talk that can completely swap linear and circular polarization from telescope input to output at some wavelengths and pointings. Nevertheless, the instrument has been successfully used for sensitive linear spectropolarimetric studies of stars as well as other studies (cf. Harrington \& Kuhn 2007, 2009a, 2009b).
	
	 Rapid modulation of the incident beam polarization, synchronous with the detector, is the one technique for removing time-dependent systematic effects. If this modulation exceeds a kilohertz, then even atmospheric seeing errors can be minimized. This technique has been developed by the solar community. For instance, the various incarnations of the ZIMPOL I and II solar imaging polarimeter have used piezo-elastic modulators or ferro-electric liquid crystals in combination with charge shuffling on a masked CCD to remove seeing induced systematic errors (Povel  2001, Stenflo et al. 1997, Gandorfer et al. 2004, Stenflo 2007). The Advanced Stokes Polarimeter (ASP) and La Palma Stokes Polarimeter (LPSP) are other notable examples (c.f. Elmore et al. 1992, Lites 1996, Mart\'{i}nez Pillet et al. 1999). This technique has been adapted for night-time spectropolarimetric use at the Dominion Astrophysical Observatory using ferro-electric liquid crystals and a fast-shuffling unmasked CCD. Though this instrument can only record a small spectral region at lower spectral resolution, fast modulation removes several systematic errors (Monin et al. in Prep).

	We present here a spectropolarimeter using retardance-varying liquid crystals. These liquid crystals add additional complexity as the retardance is chromatic and varies as a function of temperature. However, these chromatic liquid crystals can be quite useful when considering their performance characteristics. The retardance varies without physical motion of the LCVRs. This removes systematic errors caused by moving an optical element during or between exposures. Another advantage is that this type of LCVR can switch from zero to half-wave retardance in several milliseconds. These retarders can be used in any type of fast-switching application. Another trade-off lies in detector cost and complexity. In cross-dispersed echelle spectropolarimeters, the detector area must be used efficiently. If one can encode polarimetric information in a minimum number of spectral orders and minimize the required number of CCD reads, one can optimize a system for faster, more efficient operation. These trade-offs motivate the use of LCVR systems. 
	
	We show here that a Stokes-based deprojection method allows reconstruction of the incident polarization even when using LCVRs. In order to demonstrate this deprojection and improve the utility of HiVIS, we have implemented two different polarimetric modes. The first is a standard full-Stokes polarimetric mode using two rotating achromatic wave plates. The second is a fast-switching polarimetric mode using the LCVRs. With this new full-Stokes capability we were able to thoroughly calibrate and compare the polarization properties of HiVIS and quantify the sources of systematic error and polarimetric cross-talk caused by the various optical elements.
	
	In this paper we will demonstrate the calibration of the new retarders, performance characteristics of the upgraded HiVIS, and the Stokes-based deprojection routines we have developed. First, the performance and calibration of the retarders using a laboratory spectropolarimeter will be presented in section {\bf 2}. The calibration of the LCVRs as well as the Stokes-based deprojection methods will be presented in section {\bf 3}. The upgraded HiVIS spectropolarimeter properties will be presented in section {\bf 4}. Section {\bf 5} will present the HiVIS spectrograph polarization properties and section {\bf 6} will present the AEOS telescope polarization properties.

\section{Retardance Fitting and Optical Characterization}

In the Stokes formalism, the polarization state of light is denoted as a 4-vector:

\begin{equation}
{\bf I} = \left [  \begin{array}{r}
 I \\ 
 Q \\ 
 U \\
 V \\
\end{array}  \right ] 
\end{equation}

In this formalism, $I$ represents the total intensity, $Q$ and $U$ the linearly polarized intensity along position angles $0^\circ$ and $45^\circ$ in the plane perpendicular to the light beam, and $V$ the right-handed circularly polarized intensity. Note that according to this definition, linear polarization along angles $90^\circ$ and $135^\circ$ will be denoted as $-Q$ and $-U$, respectively. Furthermore, $Q$ and $U$ and $V$ are said to be orthogonal states of polarization. The degree of polarization can be defined as a ratio of polarized light to the total intensity of the beam:
	
\begin{equation}
P = \frac {\sqrt{ Q^2 + U^2 + V^2 } } {I}
\end{equation}

	The intensity of unpolarized light will be represented in this paper as:

\begin{equation}
I_{u}= (1-P) I
\end{equation}

	The normalized Stokes parameters are denoted with lower case and are defined as:

\begin{equation}
\left [  \begin{array}{r}
 q \\ 
 u \\ 
 v \\
\end{array}  \right ] 
= \frac{1}{I}
\left [  \begin{array}{r}
 Q \\ 
 U \\
 V \\
\end{array}  \right ] 
\end{equation}

	 For details on polarization of light, see an excellent text: Collett 1992. 

	 The upgrade to full Stokes polarimetry on HiVIS required the addition of an achromatic quarter-wave plate to the standard polarimetric mode and the installation of a completely new mode using two LCVRs. New calibration optics were also installed to deliver broad-band linear and circular polarization to the HiVIS spectrograph. We acquired and tested two 'compensated' LRC-200 IR-1 Meadowlark LCVRs. These retarders were designed to deliver zero to over half-wave retardance over the 6500{\AA} to 9500{\AA} range. We also acquired a Boulder Vision Optic quarter-wave plate (QWP) and half-wave plate (HWP) designed to be achromatic in the 4000{\AA} to 7000{\AA} range. Calibration polarizers consisted of a pair of Codixx wire-grid polarizers (1'' round), a pair of Edmund High-Contrast polarizers, a pair of Edmund wire-grid polarizers (50mm square) and a pair of simple sheet linear polarizers (2'' round). We will denote these as Codixx, Edmund, wire-grid and sheet polarizers respectively. 
	 
	 We describe below the characterization of these new optics using our laboratory spectropolarimeter. The spectropolarimeter is essentially a collimated halogen light source with a pair of polarizers to generate and analyze polarized light. Light is collected in a fiber-fed Ocean Optics USB 200 spectrograph. The polarimetric efficiency for all the polarizer pairs used in this paper were measured with the laboratory spectropolarimeter. The sheet, Edmund, Codixx and wire-grid polarizer pairs were first measured both crossed (minimum transmission) and parallel (maximum transmission) to characterize their degree of polarization. The wavelengths showing $\geq$95\% polarization for each polarizer pair are shown in Table 1.

\begin{table}[!h,!t,!b]
\begin{center}
\begin{small}
\label{tablams}
\caption{Optic Wavelength Ranges}
\begin{tabular}{lc}
\hline
\hline
{\bf Name} 	& {\bf Wavelength ({\AA})}	 \\ 
\hline
\hline
Sheet		& 4200-8200	\\
Edmunds 		& 5200-10000	\\
Codixx		& 5200-10000	\\
Wire-grid		& 3800-10000	\\
\hline
LCVR		& 6500-9500	\\
Wave-plate	& 4000-7000	\\
\hline
\hline
\end{tabular}
\end{small}
\end{center}
This Table shows the name of each optic used in calibrations. The top section describes the polarizers used and the measured range where the measured polarization was $\geq$95\%. The bottom section shows the design range for the retarders used. 
\end{table}	
	
	The sheet polarizer has a high degree of polarization from roughly 4800{\AA} to 8200{\AA}. The sheet polarizer was $>$80\% polarizing from the near-UV limit of the spectrograph at 3800{\AA}, above 95\% in the 4200{\AA} to 8200{\AA} region and falling sharply with wavelength after 8200{\AA}.  Both the Edmund and Codixx polarizers show high degree of polarization longward of 5200{\AA}. The wire-grid polarizers repeatably showed between 99.0\% and 99.8\% polarization for the full spectral range of 3800{\AA} to 10000{\AA}. All of the measurements were reproducible to better than 1\%. Our four well-calibrated polarizer pairs are used to test the optics in the laboratory and in the HiVIS spectrograph. In order to illustrate the repeatability and robustness of the measurements, two independent data sets were taken with different polarizers. The sheet polarizers were used for short wavelengths and the Edmund polarizers were used for long wavelengths.

\begin{figure*} [!h,!t,!b]
\begin{center}
\includegraphics[width=0.28\linewidth, angle=90]{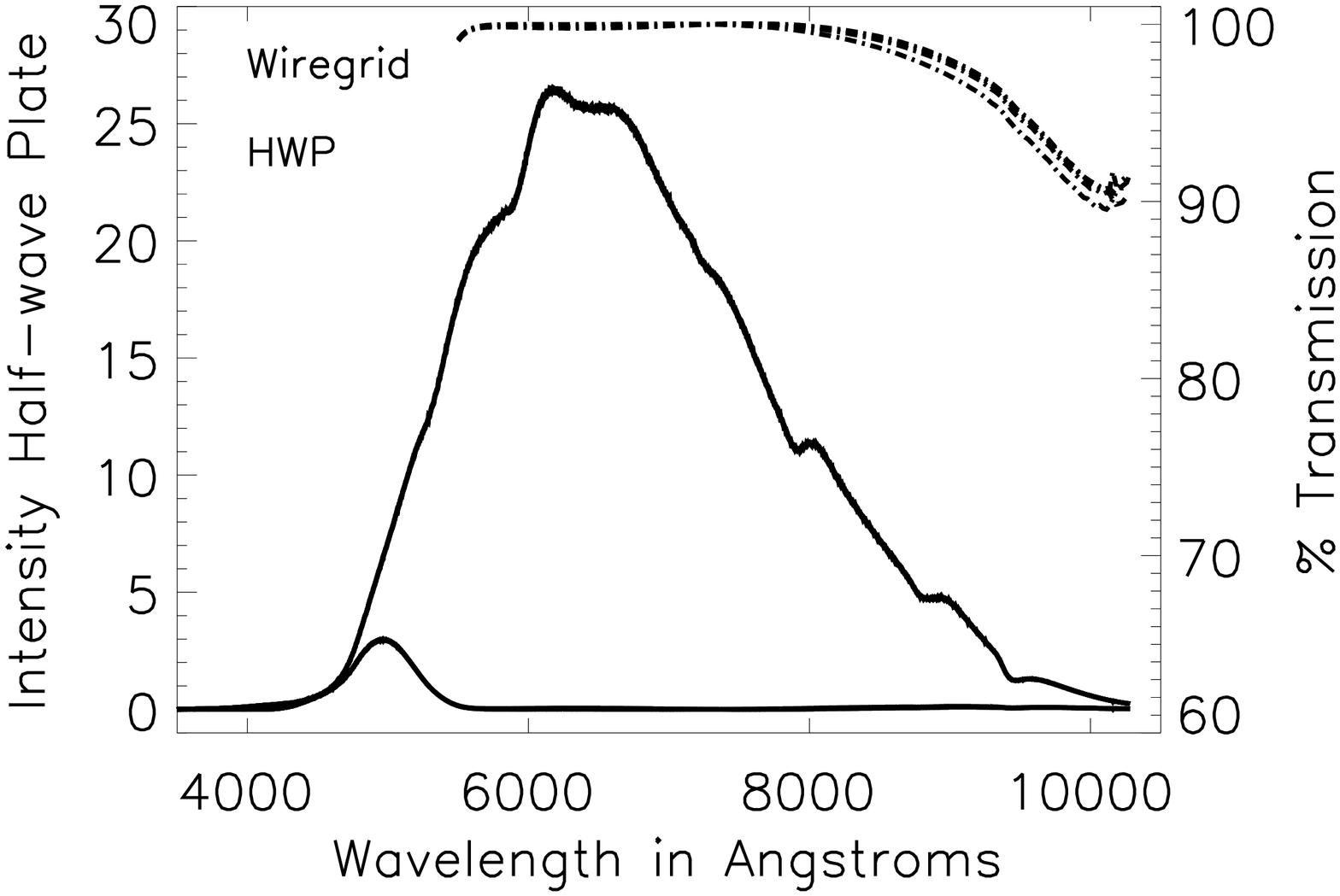}  
\includegraphics[width=0.28\linewidth, angle=90]{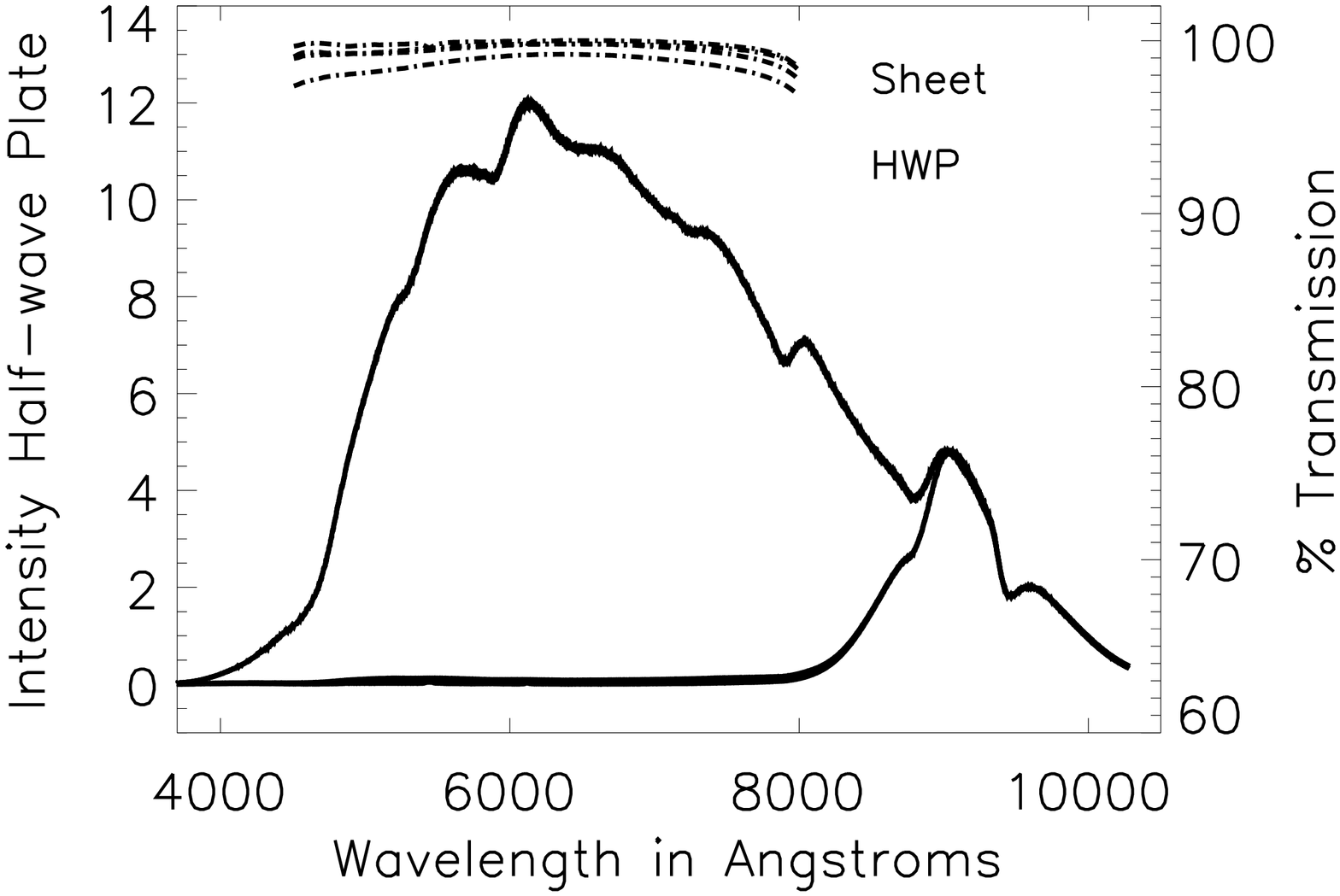}  
\includegraphics[width=0.28\linewidth, angle=90]{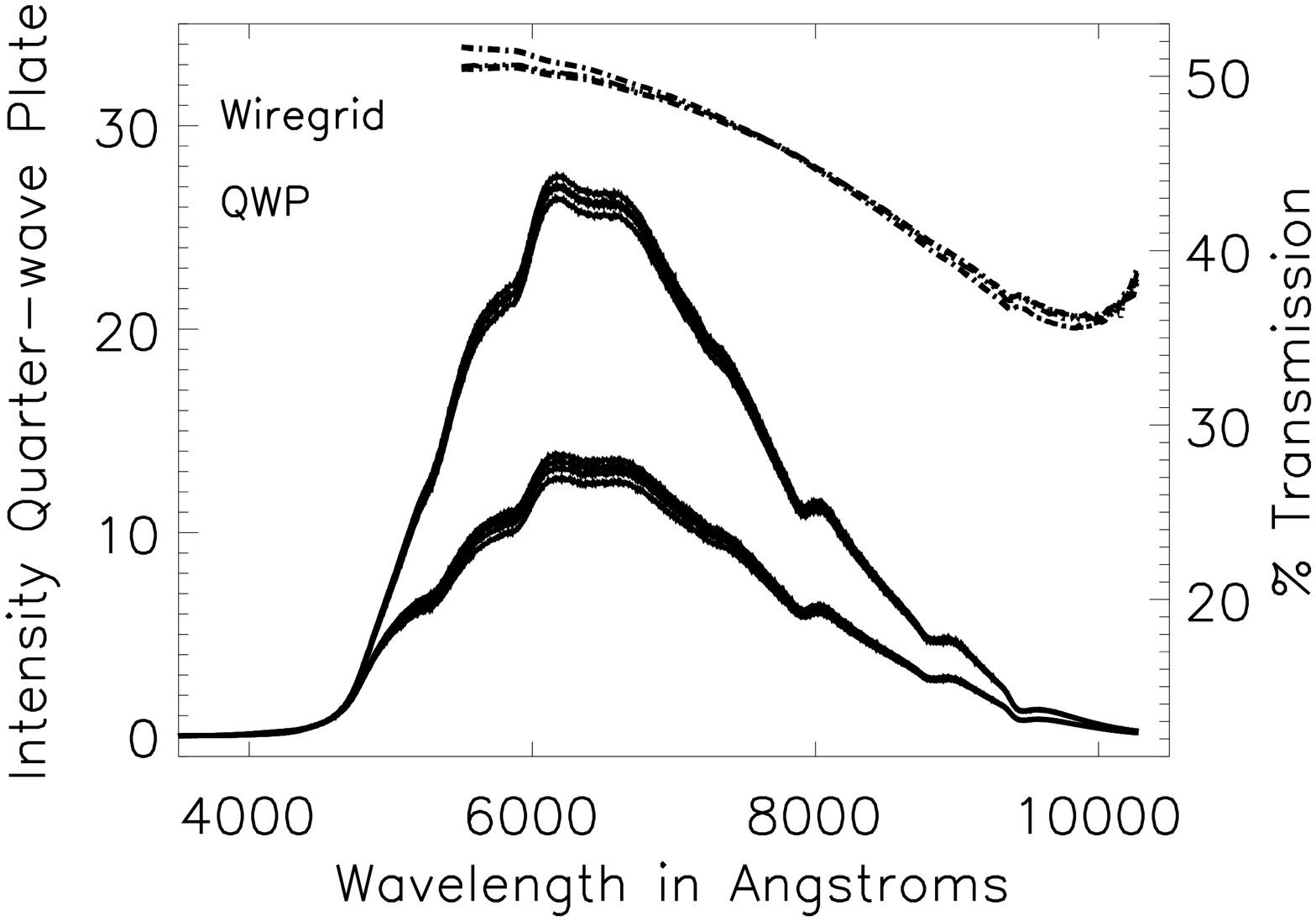}  
\includegraphics[width=0.28\linewidth, angle=90]{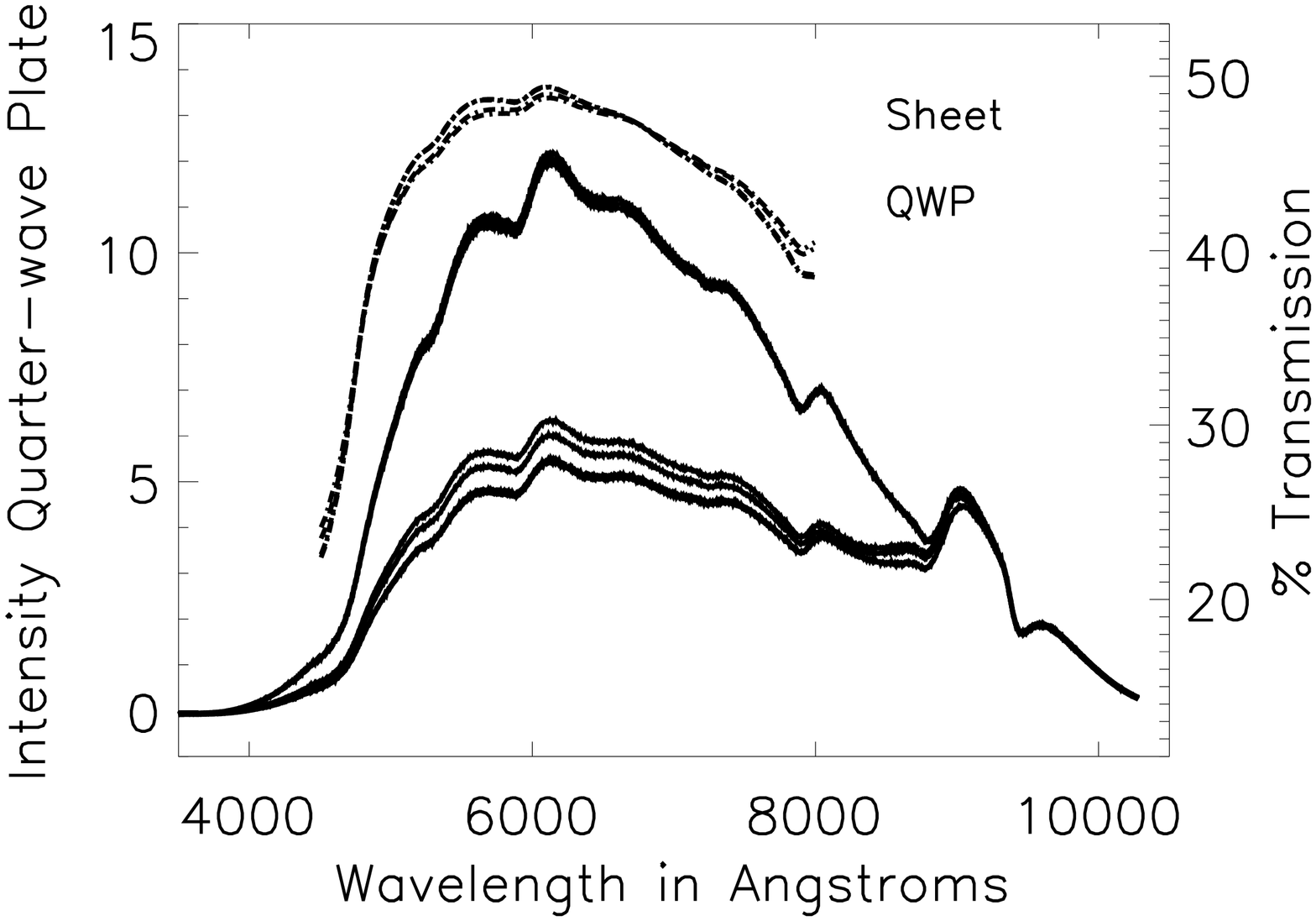}  
\caption{\label{0910-wp} This Figure shows the measured transmission spectra T($\lambda$) calculated as in Equation \ref{transeq} for the QWP and HWP (dot-dash lines). The corresponding relative intensity spectra I$_\parallel$ and I$_\perp$ are shown as solid lines. The HWP is shown in the top two panels and QWP is shown in the bottom two panels. The left hand y-axis is relative intensity. The right hand y-axis shows the transmission ratio in percent.  The relative intensities for the HWP are recorded through crossed Edmund polarizers on the top left panel and crossed sheet polarizers on the top right panel. The QWP intensities are recorded through parallel Edmund polarizers on the bottom left panel and parallel sheet polarizers on the bottom right panel. There are four individual measurements repeated at 0$^\circ$, 90$^\circ$, 180$^\circ$ and 270$^\circ$ rotation for the polarizer pairs giving rise to the many curves in each panel.}
\end{center}
\end{figure*}

\subsection{Achromatic Wave-Plate Characterization}

	To calculate retardance, we take the ratio of transmission spectra with the retarder in between polarizers. The maximum transmission spectrum $I_{\parallel}$ is recorded with parallel fast axes of both polarizers and wave plate. This yields the transmission through all optics of the system. Next, the fast axes of the retarder and second polarizer are rotated $45^{\circ}$ and $90^{\circ}$ respectively. This spectrum is labeled $I_{\perp}$, to denote the crossed polarizers. We then define the wavelength dependent transmission spectrum as
	
\begin{equation}
\label{transeq}
T(\lambda) =1- \frac{ I_{\perp}(\lambda) } {I_{\parallel}(\lambda)}
\end{equation}
	
	Thus the transmission spectrum is always between 0\% and 100\% and is independent of the spectrum of the light source. Any polarization-independent absorption caused by the retarder or polarizers is included in both I$_\perp$ and I$_\parallel$ and does not influence the transmission spectrum. Since the fast axis alignment is independent of wavelength and the polarizers are used only in their highly polarizing wavelength ranges, the transmission can be directly related to the LCVR retardance.
	
	To describe how polarized light propagates through any optical system, a Mueller matrix is constructed which describes how every incident state is transferred to an outgoing state. The Mueller matrix is a 4x4 set of transfer coefficients which multiplies the input Stokes vector to create the output Stokes vector:
	
\begin{equation}
{\bf I}_{out} = {\bf M} {\bf I}_{in}
\end{equation}	

	If the Mueller matrix for a system is known, then one inverts the matrix and deprojects a set of measurements to recover the inputs:

\begin{equation}
{\bf I}_{in} = {\bf M^{-1}} {\bf I}_{out}
\end{equation}		
	
	The inverse Mueller matrix for the system we call the deprojection matrix. One can represent the individual Mueller matrix terms as describing how one incident state transfers to another:

\begin{equation}
{\bf M} =
  \left ( \begin{array}{rrrr}
  II   &   QI   &  UI   &  VI          \\
  IQ &  QQ  &  UQ &  VQ     \\
  IU &  QU  & UU  & VU         \\
  IV &  QV  &  UV  &  VV     \\ 
  \end{array} \right ) 
\end{equation}

	In our lab spectropolarimeter $\theta$ measures the angle in the plane perpendicular to the light beam. If we define $\theta = 90^\circ$ to be the direction of Stokes Q, we can write the standard Mueller matrix for a wave plate oriented at $\theta = 45^\circ$ with its own transmission function $T_{ret}$ and a variable retardance $\phi$ as:	

\begin{equation}
{\bf M}_{ret}(45^\circ,\phi) = T_{ret} \left ( \begin{array}{rrrr}
            1 & 0 & 0 & 0 \\
            0 & \cos\phi & 0 &-\sin\phi \\
            0 & 0 & 1 & 0 \\
            0 & \sin\phi & 0 & \cos\phi 
		  \end{array} \right ).
\end{equation}

The Mueller matrix of an ideal linear polarizer looks as follows:

\begin{equation}
\label{eq:muellerpol}
{\bf M}_{pol} = \frac{1}{2} T_{pol} \left ( \begin{array}{rrrr}
            1 & \mp1 & 0 & 0 \\
            \mp1 & 1 & 0 & 0 \\
            0 & 0 & 0 & 0 \\
            0 & 0 & 0 & 0 
		  \end{array} \right ).
\end{equation}

 Where the negative sign corresponds to $\theta = 0^{\circ}$ while the positive sign is $\theta = 90^{\circ}$. Since the light of the calibration lamp is unpolarized, ${\bf I}_{lamp}=T_{lamp}\left[1,0,0,0\right]^T$, the Stokes vector of the light incident on the retarder can be written as

\begin{equation}
{\bf I}_{in} = \frac{1}{2} T_{lamp} T_{pol1}\left[1,-1,0,0\right]^T
\end{equation}

	Our analyzer (the second polarizer) has a transmission function $T_{pol2}$ and is a polarizer oriented vertically. With this Mueller matrix and input Stokes vector, we solve the transfer equation to recover the phase variations:

\begin{equation}
{\bf I}_{out} = {\bf M}_{pol} {\bf M}_{ret} {\bf I}_{in} 
\end{equation}

	If we note that $\frac{1}{2} I_{lamp} T_{pol1}T_{ret} T_{pol2} = I_{\parallel}$ we write the output Stokes vector as:

\begin{equation}
{\bf I}_{out} = I_{\parallel} \left ( \begin{array}{c}
1+cos(\phi) \\
-1-cos(\phi) \\
 0 \\ 
 0  \\
 \end{array}  \right )
\end{equation}

	This represents a purely polarized output beam with $I=-Q$. Since we are measuring this intensity through crossed polarizers, the output Stokes vector is simply I$_\perp$ as denoted above. The only variable is a wavelength dependent retardance ($\phi$). The measured output intensity becomes:

\begin{equation}
I_\perp(\lambda)=I_{\parallel}(\lambda) (1+cos(\phi)). 
\end{equation}
		 	
	In terms of the transmission function defined in Equation \ref{transeq} we get an equation for the retardance:
	
\begin{equation} \label{equphase}
	\phi(\lambda) = cos^{-1}[T(\lambda)]
\end{equation}	

	This equation applied to the measured transmission curves gives us the absolute retardance variation with wavelength ($\phi(\lambda)$). 	

\begin{figure} [!h, !t, !b]
\begin{center}
\includegraphics[width=0.8\linewidth, angle=90]{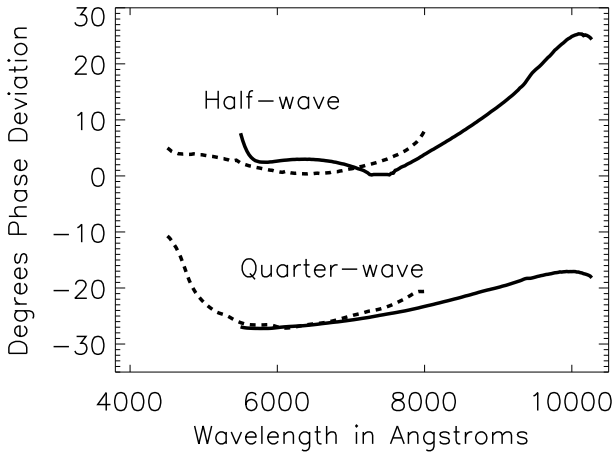}  
\caption{\label{0910-wpret} The average deviation in retardance (in degrees phase) calculated for both quarter and half wave plates through polarizer sets. The equation $\phi = cos^{-1}[T(\lambda)]$ is implemented for each observation. An arbitrary offset has been applied for clarity.}
\end{center}
\end{figure}
	
	The measurements of I$_\parallel$, I$_\perp$ and the corresponding retardances are shown in the four panels of Figure \ref{0910-wp}. The HWP measurements are shown in the top two panels. The QWP measurements are shown in the bottom two panels. Both wave plates were rotated through 360 degrees for both polarizer sets giving four independent measurements. In order to additionally test the repeatability of the mounting and aligning procedure, the QWP was unmounted and remounted for an additional set of measurements. This remounted measurement was indistinguishable from the others. 
		
	An ideal HWP oriented with the fast axis at 45$^\circ$ to the input polarization rotates the input state to the orientation of the final polarizer, giving 100\% transmission through crossed polarizers. An ideal QWP would only give 50\% transmission through both parallel and crossed polarizers. Thus, the bottom two panels have the right-hand y axis going to 50\%. Since the transmission spectra of the wave plates are only useful when the polarizers have a high degree of polarization, these transmission spectra are only plotted for long wavelengths in the left-hand boxes and are only plotted for short wavelengths in the right-hand boxes.

	Since the QWP gives 50\% transmission through both parallel and crossed polarizers, we decided to leave the polarizers parallel for the QWP measurement as it involves one less rotation of an optical element and gives less chance for systematic error. Thus the top two panels of Figure \ref{0910-wp} show measured relative intensities either near 0 or 25 while the bottom two panels show relative intensities near 12 or 25. One can also see the degree of polarization of the polarizers decreasing outside their designed wavelength ranges since the measured intensities show only small differences between crossed and parallel configurations.
			
	The corresponding retardance variation, computed as in Equation \ref{equphase}, can be seen in Figure \ref{0910-wpret}. Both QWP and HWP phase deviations measured through both sheet and Edmund polarizer pairs are shown. The HWP shows less than 10 degrees retardance deviation for the 4000{\AA} to 8000{\AA} region. The QWP is slightly less chromatic at longer wavelengths but more chromatic short of 5000{\AA}.

\section{Achromatizing an LCVR Polarimeter}
		
\begin{figure} [!h,!t,!b]
\begin{center}
\includegraphics[width=0.8\linewidth, angle=90]{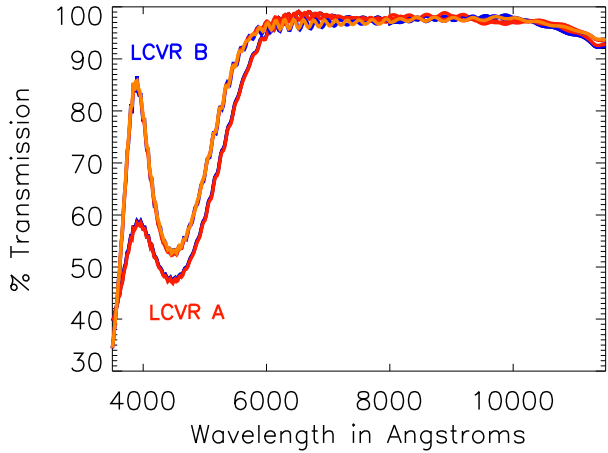}  
\caption{\label{specphot} The throughput (transmission) of the two liquid crystal variable retarders measured by the Varian Cray spectrophotometer. Each LCVR was measured at three different rotation angles and they are nearly indistinguishable on this plot. The Cray internal light source is highly polarized at most wavelengths but only a mild 'ripple' is seen at the different rotation angles.}
\end{center}
\end{figure}

	In order to use the LCVRs in a spectropolarimeter, one must know how the retardance varies with applied voltage and wavelength. The transmission function is highly chromatic but independent of polarization. Figure \ref{specphot} shows the measured LCVR transmission function. A Varian Cary 5000 spectrophotometer was used to measure the transmission of the two LCVRs we label A and B. The Cary spectrophotometer was calibrated for dark and unobscured transmission. The LCVR transmission was measured at three separate rotation angles. Since we measured the Cary light source to be highly polarized, and the LCVRs are highly birefringent and chromatic with zero applied voltage, any rotational modulation of the LCVRs could come from the polarization properties of the incident light, detector or of the LCVRs themselves.
	
	There were significant differences between the two LCVRs at short wavelengths but there was no significant dependence on LCVR orientation. Both LCVRs transmitted nearly perfectly longward of 6000{\AA}. LCVR A had a significantly lower overall transmission around 4000{\AA}. Both LCVRs showed small-amplitude ripples that were orientation dependent. LCVR A transmits over 50\% more light at very short wavelengths, but LCVR B transmits significantly more from 4000{\AA} to 6000{\AA}.

	Using the lab spectropolarimeter we measured the transmission spectra, I$_\parallel$, with the LCVR fast axis aligned with both the front and rear polarizers. The transmission spectra, I$_\perp$, was then measured for various LCVR voltages between crossed polarizers with the fast axis of the LCVR rotated by 45$^\circ$. The retardance values were recovered using Equation \ref{equphase} for each voltage in the same manner as the achromatic wave plates. It should be stressed that when the LCVR fast axis is aligned with the axis of either polarizer, there is no measurable effect with applied voltage. The LCVRs show no measurable rotation of the fast axis with voltage or wavelength.
	
	These compensated Meadowlark LCVRs are designed to have zero retardance around 5-7 volts with over half-wave retardance at low voltages. With these specifications, we convert the measured transmission spectra in to retardance spectra. We find the half and full transmission points through crossed polarizers as a function of wavelength and voltage then assign the retardance as cos$^{-1}$(T). 
	
\begin{figure} [!h, !t, !b]
\begin{center}
\includegraphics[width=0.8\linewidth, angle=90]{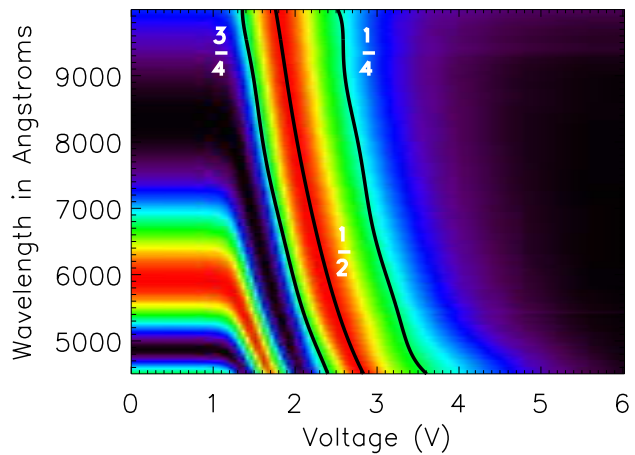}  
\caption{\label{lcvtranssurf} The LCVR-A transmission spectra through both the Edmund and sheet polarizers interpolated onto a regular voltage grid. Contours representing maximum transmission (and 1/2 wave retardance) as well as half-maximum transmission (and 1/4 \& 3/4 retardance) are overlaid. Red corresponds to maximum transmission (1) and black corresponds to zero transmission.}
\end{center}
\end{figure}

	To illustrate this process, another set of measurements of both LCVR A and B was taken with 56 separate voltages from 0V to 10V. Sampling of 0.25-0.50V per step was performed around regions of rapid change with wider sampling at low and high voltages. These measurements were in agreement with the ones presented previously. Several hundred spectra were again recorded and normalized for each voltage and averaged to create a single precise transmission spectrum. Figure \ref{lcvtranssurf} shows these transmission spectra interpolated to a regular voltage and wavelength grid. The contours for half and full transmission are overlaid representing 1/4, 1/2 and 3/4 wave retardances. 	

	 The structure of Figure \ref{lcvtranssurf} arises from the liquid crystal material properties and design of the LCVR. The liquid crystal itself is a layer of uni-axial birefringent material. The molecular axes are naturally aligned and the material is fixed between two rubbed glass surfaces to orient the material. On the inside of the glass surfaces is a thin transparent layer of indium-tin oxide that allows a voltage to be applied across the material. The material has maximal birefringence under low voltage as all the material is aligned parallel to the rubbed glass surfaces. As the voltage across the liquid crystal increases, the liquid crystal material in the center of the layer is torqued in response to the electric field. This central material aligns itself with the field while the material attached to the glass surfaces remains fixed and supplies a restoring torque. As voltage is increased the birefringence is reduced as the total degree of molecular alignment in the liquid crystal is reduced. The molecules in the center become aligned with the electric field while those at the surface remain aligned with the surface. Thus the electric field scrambles the bulk alignment of the material. Since these liquid crystals cannot sustain a DC voltage for a long time without damage, an AC square wave at 2kHz is applied which simply alternates the sign of the field while preserving the amplitude.

	One can see in Figure \ref{lcvtranssurf} that there is a region between 0 and 1 volts where there is essentially no change with voltage. This corresponds to the region where the electric field across the liquid crystal is insufficient to cause any molecular rotation in the material. There is a strong change seen from 1V to 4V as the liquid crystal molecules begin to move. Above 4 volts, the material is mostly saturated and there is essentially no birefringence and very low chromatism. 

	The two LCVRs have very similar retardance at a fixed voltage. However, two-decimal place precision in voltage is required to accurately reproduce retardances and an independent calibration must be done for each LCVR. These retardance measurements are repeatable to better than a few percent. The results of Figure \ref{lcvtranssurf} were reproduced even after a complete rebuild of the lab spectropolarimeter.

\begin{figure} [!h, !t, !b]
\begin{center}
\includegraphics[width=0.8\linewidth, angle=90]{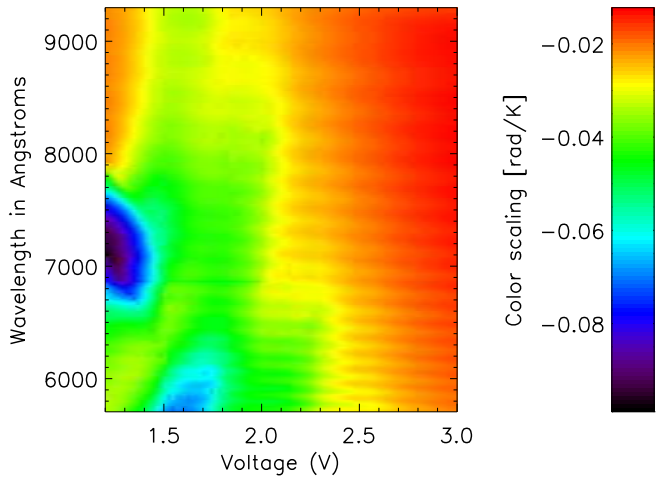}  
\caption{\label{lcvrtemp} The liquid crystal temperature dependence shown in radians induced phase change per degree temperature increase. All values are negative showing that retardance drops with increasing temperature.}
\end{center}
\end{figure}

	There is a significant temperature dependence for the LCVRs. Retardance measurements were taken at 20$^\circ$C and 40$^\circ$C. The computed retardance when heated was subtracted from the retardance measured when cool to form the temperature coefficients (in radians per degree temperature change) shown in Figure \ref{lcvrtemp}. There is a decrease in retardance with an increase of temperature ranging from 0.02 to 0.08 radians per degree. This range of voltages corresponds to the region of maximal sensitivity to LCVR input voltage, only spanning 1.2-3V. There is generally low retardance variation with temperature at higher voltages and longer wavelengths as the physical retardance values are actually lower. There is generally higher temperature sensitivity at shorter wavelengths and lower voltages as these settings correspond to higher birefringence and stronger implicit voltage dependence. This shows that temperature is an important variable that must be controlled when using LCVRs.

\subsection{Observing with an LCVR Spectropolarimeter}
	
	As a way of illustrating the deprojection process we will illustrate the extraction of the Stokes parameters from a typical observing sequence for a hypothetical spectropolarimetric observation of the 7590{\AA} TiO band. In a typical dual beam spectropolarimeter sequence, one exposure will correspond to a measurement of one Stokes parameter. There are other modulation schemes that record linear combinations of the Stokes parameters in other ways, but many night-time high-resolution spectropolarimeters function in this manner. 
	
	 In a typical sequence, a series of 6 exposures is taken corresponding to positive and negative Stokes parameters. The dual-beam analyzer gives two orthogonally polarized spectra per exposure. The orthogonally polarized exit beams become the top and bottom spectra imaged on the detector. Since the analyzer sends $I+Q$ to the top spectra and $I-Q$ to the bottom spectra and all other states are split equally between top and bottom spectra, one can simply subtract the top spectra from the bottom spectrum to obtain a polarized spectrum in one Stokes parameter. 
	 
	 The retarders act to modulate the incoming polarization so that the analyzer will produce two orthogonally polarized beams which are sensitive to other Stokes parameters in the incident beam. The typical setup with a Savart plate analyzer, shown in Figure \ref{schematic}, is to have the first LCVR with its fast axis aligned with the Savart plate axis and a retardance $\phi_1$. The second LCVR is rotated by 45$^\circ$ and has a retardance $\phi_2$. The typical procedure is to use combinations 0, 90, 180 and 270 degrees phase to achieve the modulation.
	
	The first step in this example is calibrating the liquid crystals. From measurements like those of Figure \ref{lcvtranssurf} for 7590{\AA} one finds that voltages of 6.50, 2.80, 2.10 and 1.65 correspond to the retardances of 0, 90, 180 and 270 degrees phase respectively. These voltages then become the settings for a normal sequence. The LCVRs are set to their respective phases to measure one particular Stokes parameter, say ($\phi_1$, $\phi_2$)=(0,90) for $+U$ and the exposure is taken recording the corresponding spectra. The LCVR phases for each setting can be denoted as:

\begin{equation}
\label{eqnstd}
\left ( \begin{array}{r}
 Exp 1 \\
 Exp 2 \\
 Exp 3 \\
 Exp 4 \\
 Exp 5 \\
 Exp 6 \\
 \end{array} \right ) =>
\left ( \begin{array}{r}
 \phi_1^{00}, \quad \phi_2^{000} \\
 \phi_1^{00}, \quad \phi_2^{180} \\
 \phi_1^{90}, \quad \phi_2^{090} \\
 \phi_1^{90}, \quad \phi_2^{270} \\
 \phi_1^{00}, \quad \phi_2^{090} \\
 \phi_1^{00}, \quad \phi_2^{270} \\
 \end{array} \right )=>
\left ( \begin{array}{r}
 +Q \\
 -Q \\
 +U\\
 -U \\
 +V \\
 -V \\
\end{array} \right )
\end{equation}	
	
	 This can be more compactly written as $\phi_1^{\alpha}$ and $\phi_2^{\beta}$ where $\alpha$ and $\beta$ denote the retardance angle at the nominal wavelength. The phases are denoted as $\alpha$=(0,0,90,90,0,0) and $\beta$=(0,180,90,270,90,270). Each orthogonally polarized beam exiting the Savart plate is a pure polarization state represented as a scalar intensity coefficient multiplying a Stokes vector for pure $+Q$:
	 
\begin{equation}
{\bf I}_{top} = I_{top} \left (
\begin{array}{r}
 1 \\
 +1 \\
 0 \\
  0 \\ 
 \end{array}  \right )
\end{equation}

	  What we are interested in is how the incident polarization state maps to the intensity coefficients I$_{top}$ and I$_{bot}$. This can be calculated by simply multiplying the Stokes vector of the incident light by the combined Mueller matrix for the two LCVRs at their respective orientations and retardances and a linear polarizer which represents the Savart plate analyzer. The main result is that unpolarized light is always recorded with equal intensity in both top and bottom beams. For every LCVR retardance setting of the normal sequence, the desired Stokes parameter is always recorded with a coefficient of $\pm$1 while the other Stokes parameters are recorded equally with coefficients of 0.5. The prescription for where each Stokes parameter falls in each spectrum and the two corresponding LCVR phases (shown parenthetically in degrees) is given by:

\begin{figure} [!h, !t, !b]
\begin{center}
\includegraphics[width=0.90\linewidth, height=0.4\linewidth]{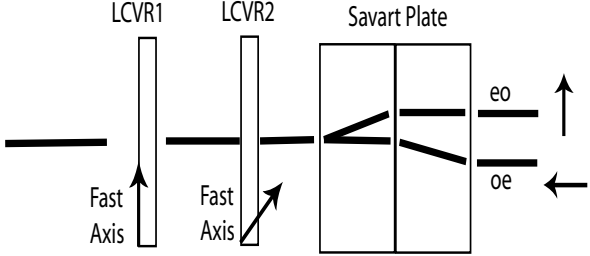}  
\caption{\label{schematic} The schematic for the HiVIS spectropolarimeter. Light incident from the left hits the first LCVR, oriented with the fast axis vertical. The second LCVR is rotated by 45$^\circ$. The Savart plate is two calcite crystals bonded such that the extraordinary beam for the first crystal becomes the ordinary beam in the second crystal. The Savart plate axis is oriented vertically, giving orthogonally polarized $\pm Q$ exit beams.}
\end{center}
\end{figure}

%\begin{equation}
\begin{align*}
I_{1top} = +Q 	+ ( I_{u}+U+V)/2  	 \quad   (\phi_1, \phi_2)=(00, 000) 	\\
I_{1bot} = -Q 	+ ( I_{u}+U+V)/2 	 \quad   (\phi_1, \phi_2)=(00, 000) 	\\
I_{2top} = -Q 	+ ( I_{u}+U+V)/2  	 \quad   (\phi_1, \phi_2)=(00, 180) 	\\
I_{2bot} = +Q 	+ ( I_{u}+U+V)/2 	 \quad   (\phi_1, \phi_2)=(00, 180) 	\\
I_{3top} = +U 	+ ( I_{u}+Q+V)/2  	 \quad   (\phi_1, \phi_2)=(90, 090) 	\\
I_{3bot} = -U 	+ ( I_{u}+Q+V)/2 	 \quad   (\phi_1, \phi_2)=(90, 090) 	\\
I_{4top} = -U 	+ ( I_{u}+Q+V)/2  	 \quad   (\phi_1, \phi_2)=(90, 270) 	\\
I_{4bot} = +U 	+ ( I_{u}+Q+V)/2 	 \quad   (\phi_1, \phi_2)=(90, 270) 	\\
I_{5top} = +V 	+ ( I_{u}+Q+U)/2  	 \quad   (\phi_1, \phi_2)=(00, 090) 	\\
I_{5bot} = -V 	+ ( I_{u}+Q+U)/2 	 \quad   (\phi_1, \phi_2)=(00, 090) 	\\
I_{6top} = -V 	+ ( I_{u}+Q+U)/2  	 \quad   (\phi_1, \phi_2)=(00, 270) 	\\ 
I_{6bot} = +V 	+ ( I_{u}+Q+U)/2 	 \quad   (\phi_1, \phi_2)=(00, 270) 	\\
\end{align*}
%\end{equation}

	This 'normal sequence' records one Stokes parameter in one exposure by encoding that parameter in a single top or bottom spectra with no sensitivity to the other Stokes parameters. For instance, Stokes $Q$ is entirely contained in the first exposure ($I_{1top}$ and $I_{1bot}$) at the optimum wavelength and the other states are equally recorded in both beams. The $I_{1top}$ spectrum minus the $I_{1bot}$ spectrum gives only Stokes $Q$ sensitivity. The sign of each state is reversed in even versus odd exposure numbers.
	
	 This sign reversal is the key to the dual beam concept as it enables systematic error subtraction and independence from detector effects. Each Stokes parameter is recorded on the same set of pixels with opposite sign. Subtraction of the two spectra removes detector noise but adds the Stokes parameters. Once this normal sequence is observed, we have recorded 12 spectra in 6 exposures where simple subtraction of individual spectra is enough to disentangle each Stokes parameter. The three exposures recording the negative Stokes parameters are used to remove CCD cosmetic effects and any pixel-to-pixel sensitivity variations. 

\begin{figure} [!h,!t,!b]
\begin{center}
\includegraphics[width=0.6\linewidth, angle=90]{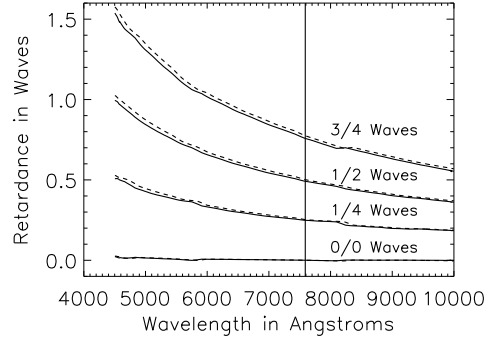}  
\caption{\label{lcvr-retardance} The retardance versus wavelength for four selected LCVR voltages. The solid line shows LCVR A and the dashed line shows LCVR B. The vertical line shows the TiO wavelength of 7590{\AA} where the LCVR retardances were optimized to be 0.00, 0.25, 0.50 and 0.75 waves.}
\end{center}
\end{figure}
		
	Even though the LCVRs are chromatic, the polarization information at all other wavelengths is still present. Extracting the polarization spectra is possible but requires a more complex deprojection process. This gives rise to wavelength-dependent sensitivities for each Stokes parameter. In order to deproject the observations, measurements of the LCVR retardance must be used. Figure \ref{lcvr-retardance} shows the retardance for the four voltages used in this sequence. There is minimal difference between individual LCVRs as the solid and dashed lines overlay quite nicely. In order to represent our two LCVR polarimeter for all wavelengths, we calculate the Mueller matrix for a 2-LCVR system with arbitrary phase. We will denote the cosine of a phase as c$\phi$ and the sine of a phase as s$\phi$. The Mueller matrix of the two LCVR system ({\bf M}$_{2LCVR}$) is just {\bf M}$_{ret}$(45,$\phi_1$) multiplied by {\bf M}$_{ret}$(0,$\phi_2$):

\begin{equation} 
{\bf M}_{2LCVR} =
\label{lcvrmm}
	\left ( \begin{array}{rrrr}
	1 & 0 & 0 & 0 \\
	0 & c\phi_2 & 0 &-s\phi_2 \\
	0 & 0 & 1 & 0 \\
	0 & s\phi_2 & 0 & c\phi_2 
	\end{array} \right ). 
	\left ( \begin{array}{rrrr}
	1 & 0 & 0 & 0 \\
	0 & 1 & 0 & 0 \\
	0 & 0 & c\phi_1 &s\phi_1 \\
	0 & 0 & -s\phi_1 & c\phi_1 
	\end{array} \right )
\end{equation}

This can be multiplied out to create the full Mueller matrix for the liquid crystals: 

\begin{equation}
{\bf M}_{2LCVR} = 
	\left ( \begin{array}{rrrr}
	1 & 0 		& 0 					& 0 			\\
	0 & c\phi_2 	& s\phi_1s\phi_2 	&-c\phi_1s\phi_2 	\\
	0 & 0 		& c\phi_1			& s\phi_1 			\\
	0 & s\phi_2 	&-s\phi_1c\phi_2	& c\phi_1c\phi_2 
	\end{array} \right )
\end{equation}

	The dual-beam spectropolarimeter essentially has two separate analyzers. The top and bottom spectra can be represented as having passed through a linear polarizer oriented vertical (0$^\circ$) and a second linear polarizer oriented horizontal (90$^\circ$). The spectra recorded through each beam can be calculated by multiplying the 2-LCVR Mueller matrix by the Mueller matrix for the linear polarizers which select the individual top and bottom beams. If the beam recorded through the vertical polarizer is the 'top' spectra and the horizontal polarization is the 'bottom' spectra, the equation for the intensity coefficients of the recorded spectra becomes:

\begin{equation}
\label{recint}
\left ( \begin{array}{r}
I_{top} \\
I_{bot} 
\end{array} \right ) = \frac{1}{2} \left (
\begin{array}{r}
 I_{u} - c\phi_2 Q - s\phi_2 s\phi_1 U + s\phi_2 c\phi_1 V  \\
 I_{u} + c\phi_2 Q + s\phi_2 s\phi_1 U - s\phi_2 c\phi_1 V 
\end{array} \right )
\end{equation}

\begin{figure} [!h,!t,!b]
\begin{center}
\includegraphics[width=0.6\linewidth, angle=90]{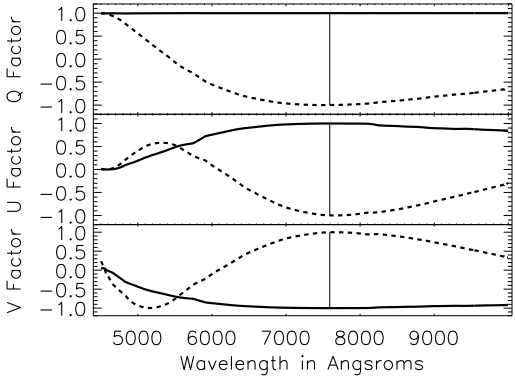}  
\caption{\label{lcvcombofactor} The LCVR efficiencies from Equation \ref{recint} as functions of wavelength using the first three retardance settings of the 'normal' sequence. The top box shows the standard '$Q$' sequence, the middle box the '$U$' sequence and the bottom box the '$V$' sequence. The solid lines are the + Stokes parameters while the dashed lines are the - Stokes parameters. Only for a narrow wavelength range around 7590{\AA} does the normal sequence properly map incident polarization to the required $\pm QUV$ as noted by the vertical lines.}
\end{center}
\end{figure}

	With the equations for the coefficients multiplying each Stokes parameter in each of the exposures, typically called the efficiencies, one can determine the chromatic effects of the LCVRs and how to deproject each of the Stokes parameters. Figure \ref{lcvcombofactor} illustrates the wavelength dependence of these efficiencies for each Stokes parameter optimized for 7590{\AA}. In the normal sequence, the top and bottom spectra record one Stokes parameter multiplied by an efficiency $\pm$1. The other Stokes parameters are all equally present in both top and bottom spectra with a coefficient of 0.5. Each panel of the Figure shows the efficiencies for an individual Stokes parameter as a function of wavelength. During the first exposure, shown in the top box, Stokes $Q$ is correctly multiplied $\pm$1 at 7590{\AA} as both LCVRs are at zero retardance and show low chromatism at high voltage. The measurement of $-Q$ (dashed curve) puts the second LCVR at half-wave retardance and lower voltage. The chromatic effects are strong enough to make both efficiencies +1 by 4500{\AA}.  Stronger chromatic effects are seen in the other Stokes parameters because of the general higher retardances (and lower voltages) present on both LCVRs.

\subsection{Retardance Fitting Deprojection}

	Given that we have now measured the retardance for the LCVRs for all possible applied voltages, an easy deprojection to attempt is simply to calculate the Mueller matrix of the LCVR polarimeter for all wavelengths using Equation \ref{lcvrmm}. We will investigate a sample observing sequence for input linear polarization as an example of this deprojection via retardance fitting.
	
	In order to properly deproject a series of 12 spectra from 6 exposures into a single polarization measurement, one must create a 'deprojection matrix'. Typically, one calculates the retardance of the LCVRs and then solves the system of equations which represent the polarized radiative transfer for the spectropolarimeter. We will illustrate this process for 7590{\AA}-optimized measurements. The observing sequence gives a set of redundant equations for the polarized spectra:

\begin{equation}
{\bf I}_{meas} = {\bf D} {\bf I}_{in}
\end{equation}	

	These equations depend on the scalar intensity coefficients and can be expressed in terms of sines and cosines of the LCVR retardances. The retardances for each LCVR ($\phi_1, \phi_2$) were expressed in Equation \ref{eqnstd}. To simplify the notation, we will number the spectra by exposure (1 to 6) and note their detector location (top or bot). We note that the retardances are implicitly functions of wavelength. The coefficients are:

\begin{equation}
\label{eqnobs}
\begin{small}
\left ( \begin{array}{r}
 I_{1top} \\
 I_{1bot} \\
 I_{2top} \\
 I_{2bot} \\
 I_{3top} \\
 I_{3bot} \\
 I_{4top} \\
 I_{4bot} \\
 I_{5top} \\
 I_{5bot} \\
 I_{6top} \\
 I_{6bot} \\
 \end{array} \right )=
 \left ( \begin{array}{cccc}
 1 	& -c\phi_2^{1}		& -s\phi_2^{1}s\phi_1^{1}		& +s\phi_2^{1}c\phi_1^{1} 	\\
 1	& +c\phi_2^{1}		& +s\phi_2^{1}s\phi_1^{1}		& -s\phi_2^{1}c\phi_1^{1} 		\\
 1	& -c\phi_2^{2} 		& -s\phi_2^{2}s\phi_1^{2} 		& +s\phi_2^{2}c\phi_1^{2} 	\\
 1	& +c\phi_2^{2} 		& +s\phi_2^{2}s\phi_1^{2} 	& -s\phi_2^{2}c\phi_1^{2}  	\\
 1	& -c\phi_2^{3} 		& -s\phi_2^{3}s\phi_1^{3} 		& +s\phi_2^{3}c\phi_1^{3} 	\\
 1	& +c\phi_2^{3} 		& +s\phi_2^{3}s\phi_1^{3} 	& -s\phi_2^{3}c\phi_1^{3} 		\\
 1	& -c\phi_2^{4} 		& -s\phi_2^{4}s\phi_1^{4} 		& +s\phi_2^{4}c\phi_1^{4} 	\\
 1	& +c\phi_2^{4} 		& +s\phi_2^{4}s\phi_1^{4} 	& -s\phi_2^{4}c\phi_1^{4} 		\\
 1	& -c\phi_2^{5} 		& -s\phi_2^{5}s\phi_1^{5} 		& +s\phi_2^{5}c\phi_1^{5} 	\\
 1	& +c\phi_2^{5} 		& +s\phi_2^{5}s\phi_1^{5} 	& -s\phi_2^{5}c\phi_1^{5} 		\\
 1	& -c\phi_2^{6} 		& -s\phi_2^{6}s\phi_1^{6} 		& +s\phi_2^{6}c\phi_1^{6} 	\\
 1	& +c\phi_2^{6} 		& +s\phi_2^{6}s\phi_1^{6} 	& -s\phi_2^{6}c\phi_1^{6} 		\\
 \end{array} \right )
\left ( \begin{array}{c}
 I_{u} \\
 Q \\
 U \\
 V \\
 \end{array} \right )
\end{small}
\end{equation}	

	In order to solve this system we simply invert the equations:
	
\begin{equation}
\label{eqn-demod}
{\bf I}_{in} = \frac{ {\bf D}^T {\bf I}_{meas} } { {\bf D}^T {\bf D}}
\end{equation}

\begin{figure} [!h,!t,!b]
\begin{center}
\includegraphics[width=0.8\linewidth, angle=90]{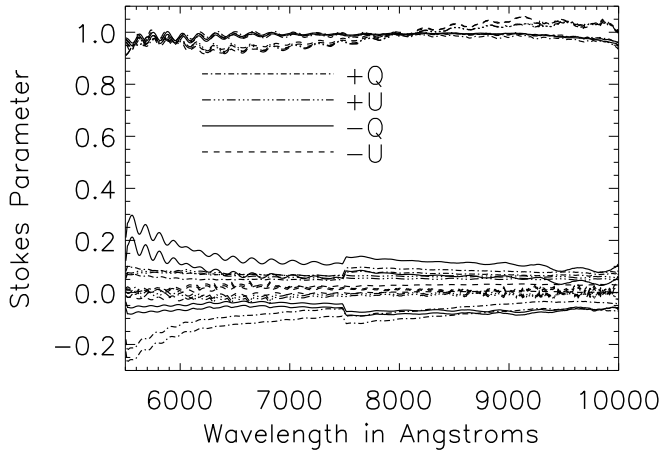}  
\caption{\label{lcvr-demod} The deprojection of pure linear polarization input using the standard deprojection of Equation \ref{eqn-demod}. Though chromatic, the deprojected outputs shown here reproduce the pure input states with values near 1. The cross-talk Stokes parameters are the curves that are below 0.3. The discontinuity around 7500{\AA} is the wavelength transition between sheet and Edmund polarizers.}
\end{center}
\end{figure}

	This is easily implemented for all wavelengths using the retardances we have calculated in the previous section. An example of the deprojection process using the lab spectropolarimeter will be shown here. The LCVR fast axes were positioned as a polarimeter with LCVR A at 0$^\circ$ and LCVR B at 45$^\circ$. Eight complete linear polarization states were input by rotating the front polarizer through 360$^\circ$ in 45$^\circ$ increments. A full polarimetric sequence was taken for each input state using the voltages chosen to optimize the polarimeter for 7590{\AA}. The overall transmission used to normalize each measurement was recorded through parallel polarizers with both LCVRs aligned with the polarizers. Since the retardance with wavelength functions are known, the complete polarimetric properties of the input can be deprojected using Equation \ref{eqn-demod}. Figure \ref{lcvr-demod} shows the deprojection for all eight input states. The input linear polarization was extracted well for all wavelengths. The input states were recovered and the cross-talk states were generally around 0.1 or less though with some chromatism and some values as high as 0.2. 

	One can see from this simple example of a normal observation sequence that there are still significant systematic errors present from phase fitting. The LCVR phases are measured in a different optical setup than when the LCVRs are used as a polarimeter. This leads to a wide range of potential errors. As we shall see in the next section, a method where phase fitting is not required performs much better.

\subsection{Stokes-Based Deprojection}

	A deprojection method which uses measurements of known input states is straightforward and direct. It incorporates all effects of the optical path and not just those of the LCVRs. There is no need to calculate the phase of the LCVRs and no interpolation is necessary. Using our lab spectropolarimeter, this is illustrated in the following example.

\begin{figure} [!h,!t,!b]
\begin{center}
\includegraphics[width=0.8\linewidth, angle=90]{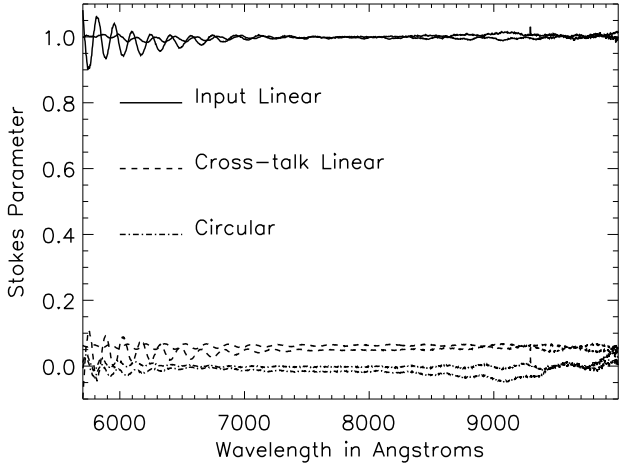}  
\caption{\label{lcvr-empir} The deprojected Stokes parameters are shown here. The input state is reproduced at a value near 1. The cross-linear and circular states are quite small, being roughly 0 for linear and 0.1 for circular. Compare this improved method with that of Figure \ref{lcvr-demod}.}
\end{center}
\end{figure}

	We inserted fully polarized light into our lab spectropolarimeter as pure inputs $+Q, +U, +V$ (as well as $-Q, -U, -V$ to show consistency). For each input state, we took the six measurements applying the `normal sequence' of voltages to our two LCVRs. In the lab spectropolarimeter we have a polarizer as an analyzer so we only record the 'top' spectra from Equation \ref{eqnobs}. Since our input space is three-dimensional, our output (prime) space will have (at most) three independent vectors, which we created by defining $Q' = I_{2top} - I_{1top}$, $U' =I_{4top} - I_{3top}$, and $V' = I_{6top} - I_{5top}$. Note that we could have chosen completely different combinations of measurements to build up our three prime space vectors, but our choice was influenced by our knowing that at 7590{\AA}, these three vectors represent the (unnormalized) values of $Q$, $U$, and $V$ from our retardance measurements. As orthonormal input basis vectors we use fully polarized light - all three pure Stokes parameters define the basis vectors. The two front linear polarizer states defining $+Q$ and $+U$ were used as well as a quarter wave plate oriented to feed in $+V$. These are our pure input states used for the Stokes-based deprojection. Then the columns of the Mueller matrix for our setup are directly given by ($Q'$,$U'$,$V'$) for each of the input basis vectors. If we now want to retrieve the input state of any measurement we again build up the quantities $I_{2top}$-$I_{1top}$, $I_{4top}$-$I_{3top}$, $I_{6top}$-$I_{5top}$, and we apply the inverse of our Mueller matrix to them. The problem turns out to be always well-posed, as one can see by looking at the (wavelength dependent) condition number of the Mueller matrix. The condition number, defined as the absolute value ratio of the highest to lowest eigenvalues of the mueller matrix, is a measure of the invertability of a linear system. In optical systems this definition is the most useful for system optimization and error propagation calculations (cf. Complain et al. 1999, DeMartino et al. 2003). The condition number is well below 10, and increasing only for wavelengths short-ward of 6000{\AA}, where the polarization performance in the system starts to drop. As two examples, Figure \ref{lcvr-empir} shows the Stokes-based deprojection method with the front polarizer input states orthogonal to the calibrations ($-Q$ and $-U$ respectively). The recovered Stokes parameters are the solid top lines near 1.0.  The dotted and dashed lines show linear and circular cross-talk.  
		
	The circular cross-talk is near zero with the linear cross-talk value being around 0.05. This represents roughly a factor of 4 improvement from the method utilizing LCVR phases presented previously. It should be pointed out that the linear cross-talk is essentially achromatic. This can be easily explained as simply an error in the rotation angle of the linear polarizer when set a negative input state. With the input state being reproduced near 1.0 and the cross-talk states being 0.05 or less, the angular error computed as 0.5 tan$^{-1}$(input / crosstalk) is less than 1.5$^\circ$. This corresponds to the error what one would expect for a manual rotation of the linear polarizer. The simple fact that the cross-talk is achromatic suggests that it is the improper rotation angle of the polarizer used to create the pure input Stokes parameters that is the cause of the error, not the improper removal of chromatic LCVR retardance.

\subsection{Benefits of Stokes-Based Deprojection}

	The main benefit of Stokes-based deprojection can be seen when considering how errors effect measured Stokes parameters. In dual-beam spectropolarimeters, the standard observing sequence is followed in order to accomplish two independent goals. One goal is to create polarization spectra that are independent of pixel-to-pixel efficiencies. Since these pixel-to-pixel variations are usually not known to better than a percent, precision spectropolarimetry requires observations to be independent of the gain. The other main goal is to choose retardances that map each Stokes parameter into these normalized double differences. For the standard sequence used with the achromatic wave plates this can be implemented as follows:
	
\begin{equation}
\label{normpol}
q= \frac{Q}{I} = \frac{1}{2}(\frac{I_{1top}-I_{2top}}{I_{1top}+I_{2top}} - \frac{I_{1bot}-I_{2bot}}{I_{1bot}+I_{2bot}})= \frac{1}{2}(q_1 + q_2)
\end{equation}
		
	One can show that this method is independent of pixel-to-pixel gains by propagating gain coefficients through Equation \ref{normpol}. Suppose that the incident radiation intensity in a top pixel is multiplied by a gain factor g$_{t}$. On a bottom pixel the gain is g$_{b}$. In the standard choice of retardances, the intensity of the incident radiation switches beams so that the I$_{1top}$ intensity before detection is the same as I$_{2bot}$. I$_{1bot}$ is identical to I$_{2top}$. Thus there are only two incident intensities denoted as I$_{t}$ and I$_{b}$ which switch places between exposure. The incident radiation is modulatd so that the gains cancel and the normalized double difference (Equation \ref{normpol}) becomes:

\begin{equation}
\label{eqgain}
q = \frac{1}{2} (\frac {g_t I_{t}-g_t I_{b}}{g_t I_{t}+g_t I_{b}} - \frac{g_b I_{b}-g_b I_{t}}{g_b I_{b}+g_b I_{t}})= \frac{ I_t - I_b }{I_t + I_b}
\end{equation}
	
	This is exactly as desired - the pixel-to-pixel gain variations have been eliminated and all that remains is a normalized difference which records one Stokes parameter. There is no systematic-error induced by pixel-dependent variations. The difference ratio method has the added benefit of removing the optical mis-alignment errors. The slit tilt and 'derivative error' caused by wavelength misalignments between 'top' and 'bottom' spectra are removed because the same pixels are used for each normalized difference (q$_1$ and q$_2$). 

	While the normalized double differences always cancel the pixel-to-pixel gains, the chromatic effects of the LCVRs require a deprojection from input to measured $QUV$. The Stokes-based deprojection utilizes calibration observations taken with as similar an optical configuration to the measurements as possible. One can see the quality of the Stokes-based deprojection by comparing the retardance-fitting deprojection of Figure \ref{lcvr-demod} with Stokes-based deprojection of Figure \ref{lcvr-empir}. Aside from the small ripples inherent in the LCVR transmission function, the Stokes-based method is much more achromatic, less prone to optical sources of error and easier to implement.

\section{HiVIS LCVR Spectropolarimeter}

	After demonstrating Stokes-based deprojection with the lab spectropolarimeter, we installed an LCVR spectropolarimeter unit and the associated calibration optics on the HiVIS spectrograph. We are presently upgrading HiVIS to include better optics, optomechanics and a fast-switching mode utilizing these LCVRs. This upgrade involves using a new charge-shifting detector. The original detectors, two 2k by 4k CCID-20 were replaced with new CCID-20s and electronics that allow charge to shuffle in one direction. If the modulation is performed faster than telescope guiding errors, a major source of systematic error can be removed. 
	
	In order to do this fast modulation, the charge will be shifted on the CCD without reading out an exposure. This charge shifting must also be synchronized with the polarimetric modulation. The spectral orders will not be allowed to cross in the charge shifting process as there is presently no shuttering during an exposure. In order to maximize signal-to-noise and make efficient use of the telescope, the number of imaged spectra must be minimized and this optimization procedure implemented. We do not need to have one exposure per Stokes parameter. As long as the polarization information is encoded in the recorded spectra in an extractable manner, the number of exposures can be reduced.

	The new HiVIS spectropolarimeter has several upgrades. The additions include a second rotating Boulder-Vision-Optic (BVO) achromatic quarter-wave plate, a remounting of all the polarizing optics with full xyz control, a new dekkar with full xyz control and polarization calibration optics. A new Savart plate was acquired with a larger clear aperture, more displacement and is more accurately bonded and a new mount which allows for easy switching between the standard wave plate mode and the fast-switching liquid-crystal mode. 
	
	The new Savart plate has a 17\% larger displacement and no detectable leak into the undesired e-e and o-o beams (compared to the 1\% measured for the old Savart plate shown in Harrington \& Kuhn 2008). The new dekkar and Savart plate allows for more sharply defined and wider spectral orders on the focal plane. The dekkar stage allows for calibration optics to be mounted just upstream of the slit and all pure input states ($QUV$) are reproduced with the achromatic wave-plates above the 95\% level for all wavelengths observed (5500{\AA}-7000{\AA}). Once this accurate reproduction of incident polarization at the slit was verified using the wave-plates the new 2-LCVR spectropolarimeter fast switching system was installed.

	We first show the properties of the HiVIS liquid-crystal system and then illustrate the Stokes-based deprojection for the system. The variations in signal-to-noise ratio with wavelength in the deprojected output will be discussed along with implications for the design and optimization of an achromatized spectropolarimeter.

\subsection{HiVIS LCVR Observations}

	In order to make the HiVIS LCVR system very robust, empirical calibration hardware and software has been developed. A new calibration unit consisting of an Edmund high-contrast polarizer and a BVO quarter-wave plate creates pure input states ($\pm Q, \pm U, \pm V$) at the spectrograph slit mirror. These pure inputs are used to derive the retardances and Mueller matrix coefficients of the LCVRs exactly as mounted and run during the night. As an example of this technique, we will use a sample set of H$_\alpha$ observations and calibrations. These observations were obtained using the Apogee 3k$^2$ detector. The Apogee has 13 spectral orders in the 5500{\AA} to 7000{\AA} range each with 3056 independent polarization measurements per order. We will define Stokes +Q as parallel to the slit.
	
	The LCVR voltages were selected to be only roughly close to the 'normal sequence' for H$_\alpha$. Deliberate rounding errors in voltage were introduced to show that, even with significant retardance selection error, the Stokes-based deprojection method works well. Voltages of 7.25, 2.95, 2.10 and 1.90 volts were chosen corresponding to retardances of -0.004, 0.215, 0.425 and 0.596 waves when mounted in the lab spectropolarimeter. This lab spectropolarimeter differs from the HiVIS setup in temperature and the two LCVRs were remounted introducing potential alignment variations. These values are rough and the deprojection does not require a derivation of the actual retardances. 

\begin{figure} [!h,!t,!b]
\begin{center}
\includegraphics[width=0.7\linewidth, height=1.1\linewidth, angle=90]{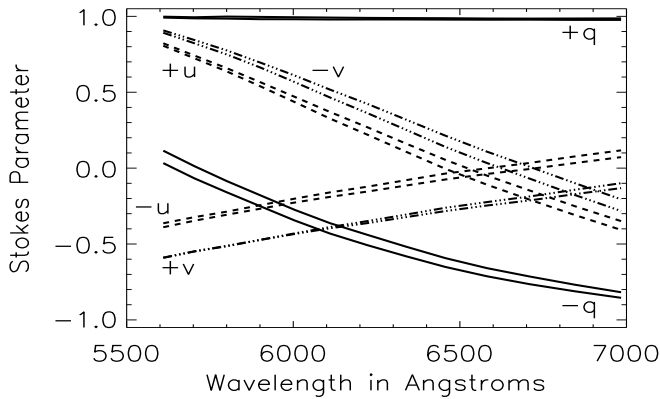}  
\caption{\label{hivis-lcvr-invpar} This Figure shows the 6 measured polarization spectra of the normal sequence. The six individual exposures produce 12 spectra which are differenced as in Equation \ref{normpol} representing $\pm QUV$. The Stokes parameters have been normalized. The solid lines show $\pm q$, the dashed line shows $\pm u$ and the dot-dash line shows $\pm v$. To illustrate the repeatability of the measurement, two complete observation sequences were taken with pure $+Q$ and $-Q$ inputs. The sign has been reversed for all measured output states for the $-Q$ input. Thus there are 24 spectra plotted here. See text for details.}
\end{center}
\end{figure}

	For each pure input state, the standard set of $\pm QUV$ observations listed in Equation \ref{eqnstd} were taken. The data were processed using the IDL reduction scripts described in Harrington \& Kuhn 2008. We found that each echelle order required using a dark subtraction from the non-illuminated regions adjacent to each spectral order. A spatially and temporally variable dark current at the 100-ADU level is present in the Apogee detector and is removed independently for each exposure. 

	If the usual dual-beam polarization analysis is performed using these observations then the input pure polarization states are poorly reproduced, as expected. The LCVR chromatism and our deliberate introduction of imperfect retardances causes significant changes in how the Stokes parameters are recorded on the detector. This is the perfect situation to illustrate the power of this deprojection technique to correct an imperfect polarimeter. Figure \ref{hivis-lcvr-invpar} shows the measured Stokes parameters derived from the six exposures for two full observing sequences. Pure $+Q$ was input for one sequence and pure $-Q$ was input for another sequence. The sign of the measured Stokes parameters has been reversed for the $-Q$ input sequence for ease of plotting. The high voltage setting corresponding to zero retardance (7.25, 7.25) shows a near-perfect reproduction of $\pm$q for all wavelengths (solid line on top) whereas the half-wave retardance on the second LCVR induces a fairly chromatic dependence with only 70\% reproduction of $\pm q$. The $u$ states are nearly zero at H$_\alpha$ but are highly chromatic in that the dashed lines cross zero but reach -.4 and +.8 at short wavelengths. The $v$ states are very similar to $u$ but with a significant offset - there is no wavelength where both + and - $v$ are zero.

\subsection{HiVIS LCVR Stokes-Based Deprojection}

\begin{figure} [!h,!t,!b]
\begin{center}
\includegraphics[width=0.75\linewidth, angle=90]{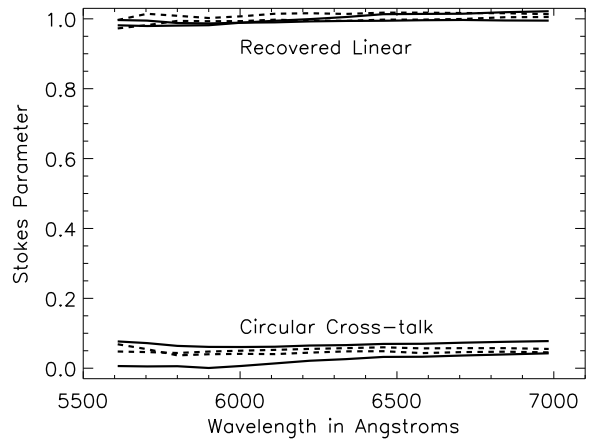}  
\caption{\label{lcvr-hi-empir} This Figure illustrates the empirical inversion process with LCVRs on HiVIS. Input states $+Q$, $+U$ and $+V$ were observed to create the deprojection matrix. The calibration optics were rotated to input negative states. These measured and deprojected negative Stokes parameters are shown here. The input state is reproduced at a value near 1 with the cross-linear and circular states are quite small. }
\end{center}
\end{figure}

	Pure input states [$+Q,+U,+V$] were observed with a full polarization sequence described above (6 exposures, 12 spectra). As each set of pure input states are observed twice, one set with lower retardances and one set with higher retardances, we actually get two independent measurements of each input state and hence two independent deprojection matrices. The measurements of each pure input state at low and high retardance form a transfer matrix at low and high retardance respectively. We then input pure negative states for testing the calibration routines [$-Q,-U,-V$]. These calibration observations also give us 6 exposures and 12 spectra which correspond to $QUV$ measurements at both low and high LCVR retardance values. The transfer matrix calculated using the positive states were inverted and used to deproject each input negative state. The resulting empirical calibration, shown in Figure \ref{lcvr-hi-empir}, shows the residuals are less than 0.1. The three input states (negative pure states) map to essentially pure measured output states (the curves near +1). The sign was reversed for clarity. The curves near zero correspond to the circular 'cross-talk' that has not been removed by this deprojection procedure. 

	These cross-talk values are quite small, being typically around 0.05 and show little chromatic variation. Considering that the calibration optics were manually rotated and are on simple post-mounts, we consider this correction to be excellent given the chromaticity of the LCVRs. As noted earlier, achromatic offsets in cross-talk can result from imprecise rotation of the calibration optics. It is the chromatic variation of the deprojected Stokes parameters that is directly caused by errors in the deprojection process. The chromatic variation in Figure \ref{lcvr-hi-empir} is much smaller than the achromatic offsets from zero.

\subsection{ HiVIS LCVR Optimization}

	Since deprojection removes the constraint that one Stokes parameter correspond to one exposure, we are free to choose retardances that give the desired precision for each Stokes parameter. One can optimize the observing sequence and retardances for a given requirement by controlling the properties of the deprojection matrix. For instance, the condition number of the deprojection matrix gives the ratio of the maximum and minimum eigenvalues (cf. Compain et al. 1999, De Martino et al. 2003). This number is an upper limit to the maximum relative variance of the corresponding deprojected Stokes parameters. Said another way: Given equal exposure times for the measured spectra, the statistical noise in the deprojected spectra will have a ratio of maximum to minimum variances that follows the condition number. We calculate the condition number by first applying the IDL single-value-decomposition routine SVDC to our deprojection matrix then taking the ratio of maximum to minimum eigenvalues ($\xi_{min}, \xi_{max}$). The condition number, c, is defined as: 

\begin{equation}
\label{eqcondnum}
c=\lvert \frac{\xi_{max}}{\xi_{min}} \rvert
\end{equation}
		
	We have investigated this relationship using our HiVIS observations. Using the deprojection matrix we have calculated for the LCVRs, we computed the condition numbers as in Equation \ref{eqcondnum}. The condition number for the low-retardance setting (high voltages) was always below 3. The condition number for the high-retardance setting (low voltages) was generally higher but always less than 4. This implies that, for a constant exposure time and statistical noise per recorded spectrum, the signal-to-noise of the resulting deprojected spectra will vary by no more than $\sqrt{3}$ and $\sqrt{4}$ respectively for the two settings. 

	We then compared the variance of our polarimetric measurements both pre- and post- deprojection. Since the Apogee detector has 13 spectral orders each with 3056 independent polarization measurements, we get 13 variance measurements in the 5500{\AA} to 7000{\AA} range. The variance $\sigma_{pre,i}$ were computed for each order $i$ before deprojection. Since the exposure times were constant, $\sigma_{pre,i}\!=\!\sigma_{pre}$. Accordingly, the variances for each order $j$ post- deprojection $\sigma_{post,j}$ were computed. Since the condition number gives the ratio of maximum to minimum variance, we define the maximum measured variance change as

\begin{equation}
\label{eqvarmax}
\delta v_{meas} = MAX_{j}( \frac { \sigma_{post,j} } { \sigma_{pre} } )
\end{equation}

	In addition, we also simulated error propagation through our calculations with simulated input Stokes parameters. In the IDL reduction package, we simply replaced the measured Stokes parameters (13 orders, 3056 wavelengths per order) with noise about zero with a variance of 1 ($\sigma_{pre}$=1). The deprojection was performed and the variances $\sigma_{post,j}^{sim}$ were determined. We can compute the variances of this simulated data in analog to Equation \ref{eqvarmax}:

\begin{equation}
\label{eqsimvarmax}
\delta v_{sim} = MAX_{j}(\sigma_{post,j}^{sim})
\end{equation}

\begin{figure} [!h,!t,!b]
\begin{center}
\includegraphics[width=0.75\linewidth, angle=90]{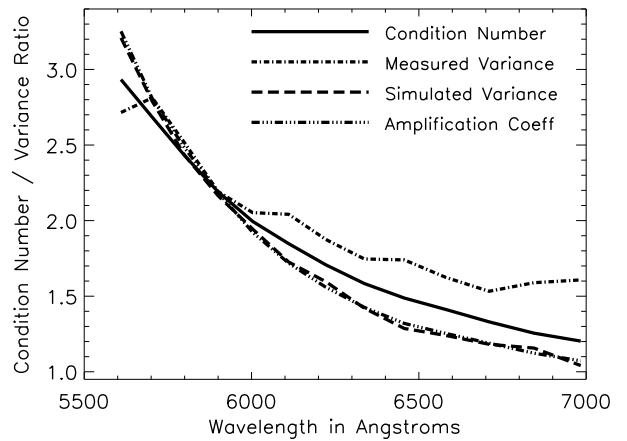}
\caption{\label{lcvr-condnum} This Figure illustrates the relationship between the condition number, measured variance, simulated variance, and 'noise-amplification-coefficient'. Since the variances are computed as ratios, all four curves should agree.}
\end{center}
\end{figure}

	Figure \ref{lcvr-condnum} shows the condition number, measured variance change ($\delta v_{meas}$) and simulated variance change ($\delta v_{sim}$) which track each other to better than 0.2 across the recorded wavelength range.
	
	The final test performed was to verify that the coefficients in the deprojection matrices agree with the relative error values. The deprojection process multiplies the inverse Mueller matrix (M$^{-1}$) by the measured polarization spectra. Each deprojected spectrum is a sum of three input spectra multiplied by the corresponding deprojection matrix element. The variance of statistical noise will increase as the square of the deprojection matrix element and the variance will also increase as the sum of these squared values. The variance should go as n=a$^2$+b$^2$+c$^2$ where a, b and c represent the row of the deprojection matrix (M$^{-1}$). This we term the 'noise amplification coefficient'. This coefficient also agrees with the simulated variance and the condition number to better than 0.2 as seen in Figure \ref{lcvr-condnum}. If we define each term of the deprojection matrix as M$^{-1}_{ij}$ for the i$^{th}$ row and j$^{th}$ column, we can define these noise amplification coefficients n$_i$ as:

\begin{equation}
n_i = \sum_j (M^{-1}_{ij})^2
\end{equation}

	Since these noise amplification coefficients are a direct measure of the signal-to-noise computed for each deprojected Stokes parameter, they can be used to optimize an observing sequence for any desired outcome.

	There are many schemes for tuning polarimetric measurements where the properties of the retarders (fast-axis orientation and retardance) are varied in order to give some specified performance of the condition numbers. There are efficiency-balanced schemes, which maintain condition-numbers of 1, optimize for equal sensitivity between linear ($QU$) and circular ($V$) or optimize on other constraints. (c.f. Nagaraju et al. 2007, Complain et al. 1999, DeMartino et al. 2003). 
	
	The ultimate limitations of this system lie in the measurement of the Mueller matrix terms. In this example, even where manual rotation of two different optics induced a substantial systematic error, the achromatized performance was acceptable. The measured cross-talk error in the output Stokes parameters was satisfactorily achromatic. Uncertainties in voltages or calculated LCVR phases do not influence the performance of the Stokes-based deprojection. There were systematic offsets that can be explained as 1-2$^\circ$ errors in rotation of the calibration optics. The chromatic error in the recovered polarization are significantly lower. A motorized stage for the calibration optics should improve the achromatization even further.

	This system is designed to be used for line spectropolarimetry which is an inherently differential technique. The continuum polarization accuracy of this system is limited by these deprojection uncertainties, as well as the polarization induced by the optics of the system. However, this calibration is quite sufficient to create a stable polarimetric reference frame for measurements of polarization across individual spectral lines. One of the most widely used high-resolution spectropolarimeters, ESPaDOnS, has a cross-talk of 2\% and a wavelength-dependent, time-variable continuum polarization (Donati et al. 2006, ESPaDOnS Instrument Website \footnote{www.cfht.hawaii.edu/Instruments/Spectroscopy/Espadons/}). Absolute polarization measurements suffer from many systematic errors but the change in instrument properties over an individual spectral line are negligible. Hence, achromatization of the instrument performance to an accuracy of a few percent is quite sufficient to allow for accurate differential measurements.

\section{Polarization Properties of the HiVIS Spectrograph}

	In order to perform accurate polarimetry of astronomical sources, a polarization calibration of the telescope and spectrograph optics is also required. The deprojection procedure we have just outlined is quite general and can also be applied to undo cross-talk induced by other optical elements in an instrument. The AEOS telescope with 5 oblique fold mirrors and the HiVIS spectrograph with its image rotator is a suitable example. The AEOS telescope induces pointing-dependent cross talk. The image rotator in HiVIS induces rotation-dependent cross-talk. The spectrograph itself also induces cross-talk even without the image rotator. Since the Savart plate analyzer is just after the slit, only the optics before the slit induce polarimetric effects.  In this section, we will use the achromatic wave-plate spectropolarimeter to measure the polarization properties of the HiVIS spectrograph optics. 
		
	The image rotator mount was aligned and remounted to allow repeatable removal and re-insertion of the image rotator with negligible influence on the spectrograph beam path. The optical path from the last coud\'{e} mirror through the spectrograph fore-optics to the analyzer has many elements. There are three near-normal incidence fold mirrors and a 5.5m focal length collimator before coming to the image rotator (3 oblique reflections). Another fold, the tip-tilt mirror, a 1.1m focal length sphere and another fold are used to create the stellar image at the slit. To calibrate with pure linear polarization, the wire grid polarizer was mounted at the calibration stage inside a mask to polarize the diffusion screen and to block unpolarized light. This serves as a polarized flat field creating pure $QU$ inputs to the spectrograph fore-optics.

\begin{figure} [!h,!t,!b]
\begin{center}
\includegraphics[width=0.75\linewidth, angle=90]{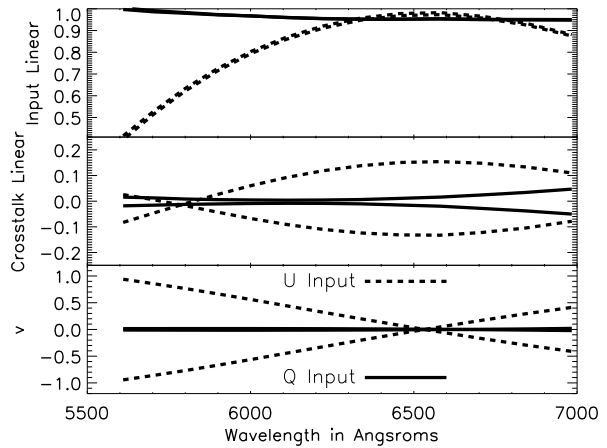}  
\caption{\label{imrot-stokespar00} This Figure shows the output Stokes parameters for four inputs ($+Q, +U, -Q, -U$) with the image rotator in the nominal orientation. Solid lines are for $\pm Q$ inputs, dashed lines are for $\pm U$ inputs. The top box shows the measured output state for the same input state (eg. $+Q$ out for $+Q$ in). The middle panel shows the non-input linear Stokes parameter showing linear cross-talk of up to 0.2. The bottom panel shows circular polarization and is entirely cross-talk.}
\end{center}
\end{figure}

\subsection{Polarization With Image Rotator}
		
	The polarization response of the spectrograph fore-optics was first measured with the image rotator in the beam and aligned vertically. Pure $\pm Q$ and $\pm U$ states were input. The cross-talk effects can be best illustrated by separating the output measurements into three components - the reproduction of the input state, the linear cross-talk and the circular-cross-talk. The measured output separated in this manner is shown in Figure \ref{imrot-stokespar00}. Only input $U$ shows significant wavelength-dependent cross-talk. The overall degree of polarization, p, calculated as p$^2$=$q^2+u^2+v^2$, is very high for both $Q$ and $U$ inputs. The values range from 96-100\% and thus the 'depolarization' terms are small.
	
	The negative and positive input states give results that are opposite in sign but very similar in magnitude and morphology. The linear cross-talk of the middle panel shows symmetry about zero and a very small 0.1-0.2 amplitude. There is significant linear cross-talk for input $U$ and a small amount for input $Q$ at longer wavelengths. The third panel shows high $UV$ cross-talk for input $U$ at both short and long wavelengths with essentially a linear dependence with wavelength. The $UV$ cross-talk fortuitously goes to zero around 6500{\AA}.
	
	The induced polarization is measured to be quite small, being under 5\% to 6\% in these wavelengths. Since the induced polarization and depolarization terms are small with nearly 100\% reproduction of the input degree of polarization, the Mueller matrix for the HiVIS system is essentially a 3x3 matrix involving $QUV$ terms in rotation and linear-circular cross-talk. 
	
	To investigate the image rotator further, the polarization properties of the spectrograph are measured with the image rotator at angles of -90,-45,0,+45 and +90 degrees. Induced polarization (unpolarized flat field measurements) as well as purely polarized inputs were measured at four polarizer orientations ($+Q, +U, -Q, -U$) for each of the 5 image rotator orientations. This set of 20 observations allows us to characterize the wavelength-dependent changes with rotator angle. 
	
	The first thing to note about the flat field is that there is significant linear polarization for all the image rotator orientations. The $q$ terms all generally decrease to longer wavelengths while the $u$ terms are more generally flat with wavelength. The $\pm$90$^\circ$ observations show a very significant $u$ of 0.1-0.15 though with low $v$. The $\pm$45$^\circ$ observations show significant $v$ at short and long wavelengths, with all observations showing zero induced circular polarization around 6200{\AA}, 300{\AA} shorter than other orientations. This shows that the induced polarization is at the 2\% to 15\% level and is highly dependent on the image rotator. 
	
	The overall polarization properties of Figure \ref{imrot-stokespar00} are generally reproduced at the varying image rotator orientations. The polarization reference frame rotates with the image rotator as expected. There is some change in the exact form of the linear to circular cross-talk but the relative amplitude does not change by much. The linear cross-talk is roughly double and there is some additional $QV$ circular cross-talk when the rotator is oriented $\pm$45$^\circ$. We should note that some calibrations were repeated with the polarizer at different spatial locations and the resulting observations did not change significantly.

\subsection{Polarization Without Image Rotator}

	Given the strong polarization variation when using the image rotator as well as the high fold angles of the mirrors, we decided to implement a mode with the image rotator removed from the beam. This removes our control of the projected slit orientation on the sky. However, we gain some very positive benefits. The removal of moving oblique-fold mirrors greatly reduces the complexity, magnitude and wavelength-depenence of the cross-talk. We also gain in throughput and get reduced static wavefront errors caused by the high-angle folds. The AEOS telescope itself has five 45$^\circ$ folds before reaching our spectrograph, but now there are no high-angle folds in the spectrograph optics. Though the rotator adds roughly 15cm to the optical path after the collimator, this is a small change given the 550cm path length from the collimator to the tip tilt pupil image. The flat field illumination pattern did not significantly change and the imaged spectral orders are essentially identical with and without the image rotator. The alignment of the system is good enough that the image rotator can simply be removed and replaced as needed without changing any other mirror orientations. 

\begin{figure} [!h,!t,!b]
\begin{center}
\includegraphics[width=0.7\linewidth, height=1.05\linewidth, angle=90]{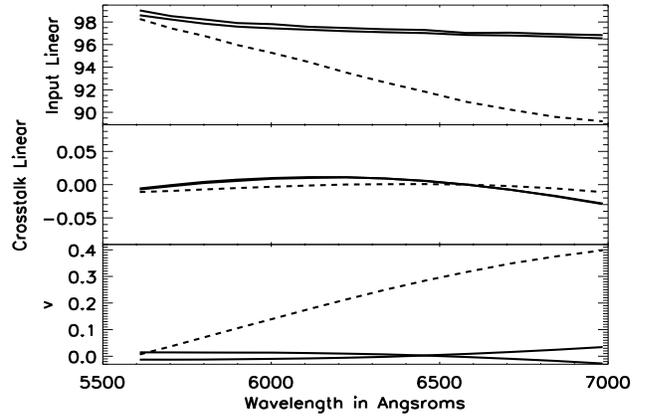} 
\caption{\label{noimrot-pol} This Figure shows the measured polarization properties with the image rotator removed. Pure input $Q$ is shown with a solid line and was measured twice. Pure input $U$ is the dashed line. The top panel shows the reproduced linear polarization input ($Q_{in}$ to $Q_{out}$, $U_{in}$ to $U_{out}$). The middle panel shows linear cross-talk. The bottom panel shows the measured Stokes $V$ running from 0 to about 0.4. }
\end{center}
\end{figure}
	
	The polarization measurements showed that the fore-optics induced polarization is typically 5\% to 6\% percent depending on the illumination pattern. The flat field screen and tip-tilt guiding piezo were moved to measure the response to a change in the illumination pattern. There are differences caused by the varying illumination, but a coherent pattern is easily recognized. The flat field is dominated by Stokes $q$ being around 0.05 to 0.06.  Stokes $u$ is 5 to 10 times smaller, being 0.006 to 0.012. The induced circular polarization, $v$, is another order of magnitude smaller than this. The total degree of polarization is essentially the degree of linear polarization of 5\% to 6\% and this is essentially all Stokes $q$. This is quite a small variation given the 2\% to 15\% that can be induced with the image rotator at various orientations.
	
	The major improvements can be seen in the reproduction of pure input linear polarization. Figure \ref{noimrot-pol} shows the measured polarization properties for input $Q$ (repeated twice, solid lines) and U (dashed lines). The top panel shows the reproduced linear polarization input at a roughly constant 98\% for input $Q$ and a 98\% falling to 90\% for an input $U$. The middle panel shows the linear cross-talk. The chromatic variation is very small and corresponds to less than 1$^\circ$. The bottom panel shows the measured Stokes $V$ which is nearly 0 for input $Q$ (just as with the image rotator in place) but only 0 rising to 0.4 for an input $U$ (as opposed to $\pm$1 changing to $\mp$1 with the image rotator in place). The general depolarization is much improved. This can be seen as the pure linear inputs being reproduced at the 90\% to 98\% level. The total degree of polarization is always above the 98\% level. The mirror-induced rotation of the plane of polarization is nearly negligible, in contrast to the 10$^\circ$ seen with the image rotator. The induced $V$ was very strong and very chromatic with the image rotator, but the induced $V$ now has a maximum of 0.4 only for input $U$ and only at long wavelengths. There is essentially no circular cross-talk at 5500{\AA}.

	As we had seen earlier, both the induced polarization and the depolarization are small. This alone suggests that one can assume that any linear to circular cross-talk has a corresponding circular to linear component. In order to verify this assumption, we remounted the slit calibration optics (high-contrast polarizer and quarter-wave plate) just down stream of the diffusion screen in order to create a Stokes $V$ input. We obtained several full-Stokes observations with both $+V$ and $-V$ inputs at various combinations of polarizer and quarter-wave plate orientations. The input $V$ had $U$ cross-talk in direct response to the $UV$ cross-talk seen in Figure \ref{noimrot-pol}. We also measured linear polarization as a consistency check and found excellent agreement.

	As expected, the image rotator is the most significant source of induced polarization and cross-talk in our polarimeter. Removing the image rotator created several polarimetric advantages. The circular polarization cross-talk in the system is $\pm$1 linearly changing to $\mp$1 from 5500{\AA} to 7000{\AA} for linear polarization input at 45$^\circ$. This induced circular polarization becomes much smaller without the image rotator, showing only $UV$ cross-talk rising to 0.4 at long wavelengths. The chromatic rotation of the plane of polarization was up to 10$^\circ$ with the rotator but is less than 1$^\circ$ without the rotator. In general the depolarization is low both with and without the image rotator. The induced polarization without the rotator is almost entirely Stokes $q$ at 5\% to 6\% with negligible induced circular polarization and with only a moderate variation with changing illumination patterns. This is to be expected as Stokes Q is defined parallel to the slit which is parallel to the image rotator axis. The spectrograph is more easily deprojected without the rotator as the condition number of the deprojection matrix is much smaller. This allows us to recover an accurate polarimetric reference frame at the entrance to the spectrograph and to separate the telescope polarization properties from the spectrograph.

\section{AEOS Telescope Polarization}

	Knowing the polarization properties of the HiVIS spectrograph with and without the image rotator, we can now quantify the polarization properties of the AEOS coud\'{e} path. A simple experiment to illustrate the cross-talk is to look at twilight at the zenith. The linear polarization properties of twilight observed with HiVIS were investigated in Harrington \& Kuhn 2008. They adopt a simple Rayleigh scattering model (cf. Liu \& Voss 1997, Lee 1998, Cronin et al. 2006). With the Sun on the horizon in the west the zenith is highly polarized with the polarization angle pointing north-south. This is a simple approximation that holds quite generally (cf. Pomozi et al. 2001, Suhai \& Horvath 2004, Horv\'{a}th et al. 2002). 
	
	The AEOS coud\'{e} path has five 45$^\circ$ reflections and there is a pair of mirrors corresponding to the azimuth axis and a pair of mirrors corresponding to the altitude axis. If we point the telescope at the zenith and simply change azimuth through 360$^\circ$, we can measure the polarization properties of this coud\'{e} path as two mirrors cross and uncross. In Harrington \& Kuhn 2008, the linear polarization of the zenith was found to decrease from 70\% to 80\% at some pointings to 15\% at other pointings. We now know that without the image rotator there is nearly no linear-circular cross-talk around 5500{\AA} but there is 0.4 at 7000{\AA}. 
	
\begin{figure*} [!h,!t,!b]
\begin{center}
\includegraphics[width=0.35\linewidth, height=0.45\linewidth, angle=90]{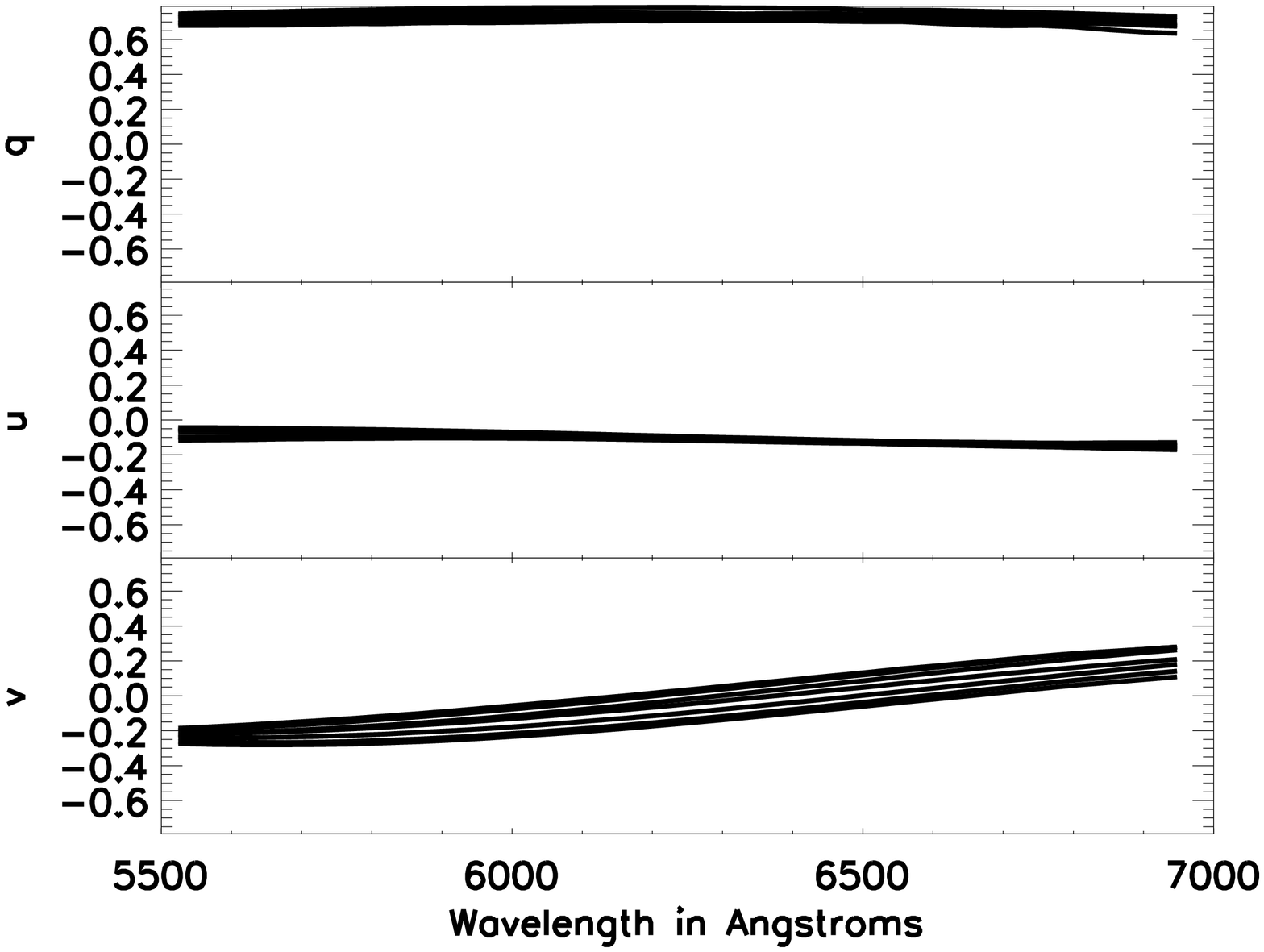} 
\includegraphics[width=0.35\linewidth, height=0.45\linewidth, angle=90]{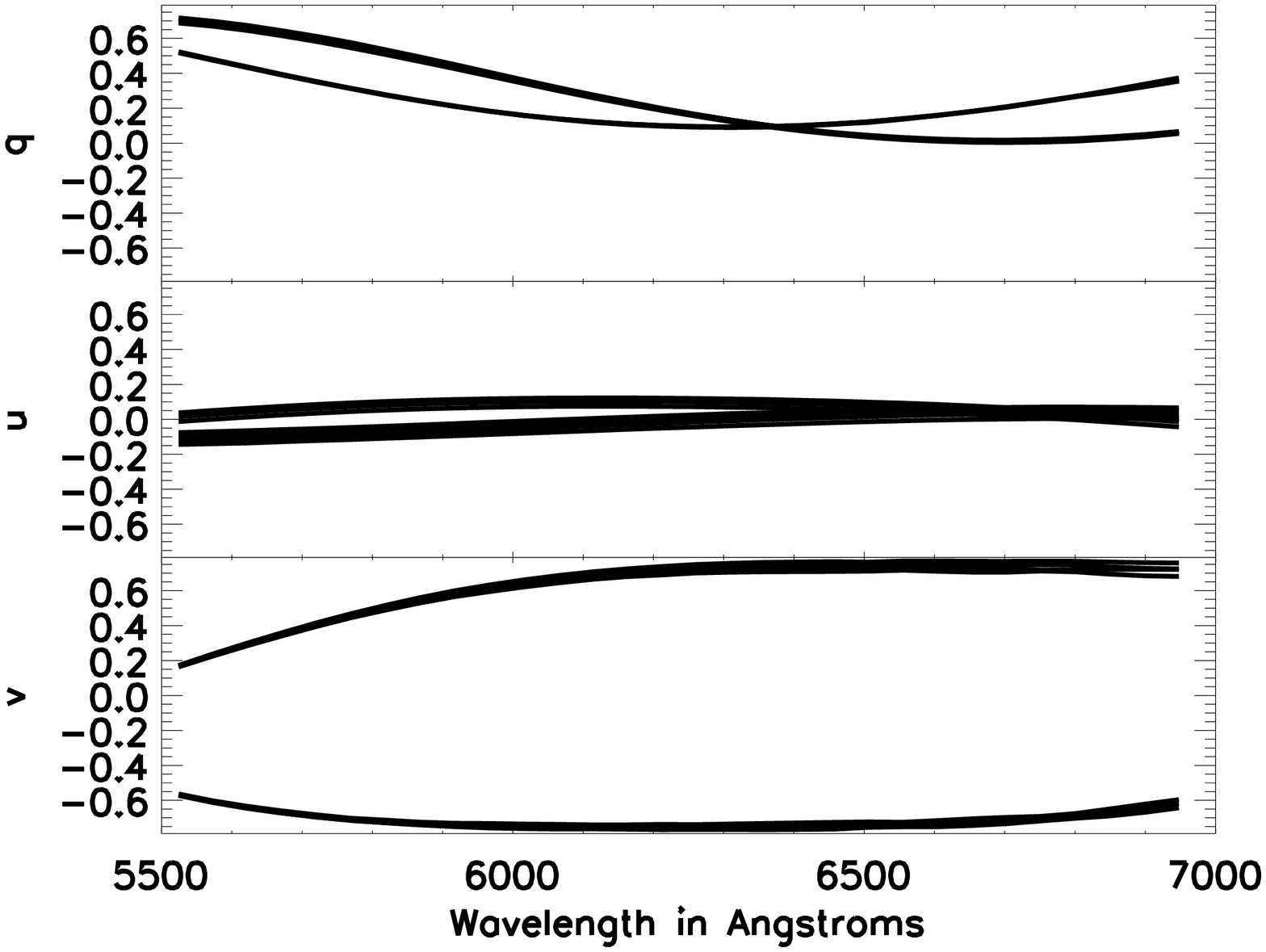} 
\caption{\label{twilight-pol} This Figure shows the measured twilight polarization properties using HiVIS without the image rotator. Each panel shows Stokes $quv$ for twilight pointing at the zenith. The left panel shows cardinal pointings: north, east, south and west. The right panel shows non-cardinal pointings: north-east, south-east, south-west, north-west. The degree of polarization for all measurements was always 70\% to 80\%. At non-cardinal pointings, the linear-circular cross-talk was severe and reached 100\% at 6500{\AA}.}
\end{center}
\end{figure*}

	We repeated the twilight experiments without the image rotator using the new full-Stokes capabilities with achromatic wave plates. Figure \ref{twilight-pol} shows the measured Stokes parameters of twilight at the zenith on September 5th 2009. On this night the Sun set at 280$^\circ$ azimuth. The left hand panel shows the Stokes parameters at azimuths of north, east, south and west (10, 100, 190 and 280). The right panel shows north-east, south-east, south-west and north-west (55, 145, 235 and 325). One can see that for the cardinal direction pointings of the left panel, the linear polarization is almost all Stokes $q$ with $v$ varying from -0.2 to +0.2 from 5500{\AA} to 7000{\AA}. Conversely, at the non-cardinal pointings of the right panel, the observed polarization is substantially more chromatic and is entirely Stokes $v$ at some wavelengths. The degree of polarization is again observed to be always 70\% to 80\% for all wavelengths. With these new observations we can conclude that the loss of linear polarization and the rotation of the plane of polarization seen in Harrington \& Kuhn at 6500{\AA} is due to linear-circular cross talk. We are in the process of measuring the telescope polarization properties at all pointings.

\section{Discussion And Conclusions}

	We have presented a Stokes-based deprojection method applied under various circumstances to highly chromatic astronomical systems. These methods are an easy and robust way to restore a reasonable polarization reference frame to a system with very significant chromatism and cross-talk. Since it is often more cost-effective to add spectropolarimetric capabilities to existing instruments, one typically has to deal with substantial systematic effects. Stokes-based deprojection is a straightforward and repeatable method to remove some of these errors. 
	
	A simple laboratory spectropolarimeter was built to characterize and test our new optics as well as to demonstrate the Stokes-based deprojection methods in the lab. Retardances of our achromatic wave plates as well as LCVRs were measured. We verified the achromatic performance of the wave plates over their designed wavelength range. The LCVRs were characterized by creating a grid of retardances versus wavelength and voltage. An LCVR spectropolarimeter was constructed and a standard sequence of observations were taken. The measurements were first deprojected using standard retardance-fitting. This retardance based deprojection was then compared to the Stokes-based deprojection method. The Stokes-based method was both easier to implement, less chromatic and more accurate.
	
	The Stokes-based deprojection methods were then demonstrated with a new HiVIS LCVR spectropolarimeter. HiVIS now includes several optical / mechanical improvements, a polarization calibration stage, a new full-Stokes achromatic wave plate mode and a liquid-crystal mode. This LCVR mode is now fully characterized to allow the choice of polarization sensitivities optimized for a wide range of applications. We have shown that one need not map individual Stokes parameters to individual exposures. For instance, we are implementing a mode where our new fast shuffling detector can interleave orders on the focal plane. With this information, we can implement a mode where the condition number of the deprojection matrix is optimized for different sensitivities within each exposure. Linear, circular or full-Stokes sensitivity can be specified within each frame to minimize systematic error in measurements of each Stokes parameter.
		
	We then used the achromatic wave-plate mode to fully calibrate the HiVIS spectrograph and show, as expected, that the main polarimetric effect of the spectrograph mirrors is linear-circular cross-talk caused by the image rotator. With this image rotator removed, the spectrograph still has linear-circular cross talk at longer wavelengths at the 0.4 level. We then used this spectrograph calibration information to show that the AEOS telescope also induces 100\% linear-circular cross-talk at the zenith when at non-cardinal azimuths. The image rotator and telescope can both induce 100\% linear-circular cross-talk. These polarimetric results are repeatable and as such, may be calibrated. Accurate absolute polarimetry from any modern alt-az telescope requires careful cross-talk calibration as we describe here.
	
	Stokes-based deprojection applied to highly chromatic systems will significantly improve the repeatability and accuracy of spectropolarimetric measurements. The LCVR system, though nominally chromatic, can be used to obtain broad-band spectropolarimetry with good accuracy and repeatability. This achromatization allows one to realize the performance advantages of a fast-switching system with no moving parts. We outlined how systematic errors resulting from rotating wave-plates and slow modulation may now be eliminated from the HiVIS spectropolarimeter.

\acknowledgements

	This program was partially supported by the NSF AST-0123390 grant, the University of Hawaii, the AirForce Research Labs (AFRL) and an SNF grant PE002-104552.


\begin{thebibliography}{}
\bibitem[Beck 	(2005a)] 	{bec05a}		Beck C. et al., 2005a, A\&A, 437, 1159
\bibitem[Beck 	(2005b)] 	{bec05b}		Beck C. et al., 2005b, A\&A, 443, 1047
\bibitem[Comp 	(1999)]	{com99}		Compain E. et al., 1999, ApOpt, 38, 3490
\bibitem[Collet	(1992)]	{col98}		Collett E., {\it Polarized Light: Fundamentals and Applications}, 1992, CRC
\bibitem[Cron  	(2006)]	{cro06}		Cronin T.W. et al., 2006, ApOpt, 45, 5582
\bibitem[DeMa	(2003)] 	{dem03}		DeMartino A. et al., 2003, Optics Lett., 28, 616
\bibitem[Dona	(1999)] 	{don99}		Donati J.F. et al., 1999, ApJS, 134, 149
\bibitem[Elm 	(1992)]	{end92}		Elmore D.F. et al., 1992, SPIE, 1746, 22
\bibitem[Evers	(1998)]	{eve98}		Eversberg T. et al., 1998, PASP, 110, 1356
\bibitem[Gand	(2004)]	{gan04}		Gandorfer A.M. et al., 2004, A\&A, 422, 703
\bibitem[Harri	(2006)]	{har06}      	Harrington D.M. et al., 2006, PASP, 118, 845
\bibitem[Harri	(2007)]	{har07}   		Harrington D.M. \& Kuhn J.R., 2007, ApJL, 667, L89
\bibitem[Harri	(2008)]	{har08}    	 	Harrington D.M. \& Kuhn J.R., 2008, PASP, 120, 89
\bibitem[Harri	(2009a)]	{har09a}  		Harrington D.M. \& Kuhn J.R., 2009a ApJS, 180, 138
\bibitem[Harri	(2009b)]	{har09b}  		Harrington D.M. \& Kuhn J.R., 2009b ApJ, 695, 238
\bibitem[Horv	(2002)]	{hor02}		Horv\'{a}th et al., 2002, ApOpt, 41, 543
\bibitem[Ichim	(2008)]	{ich08}		Ichimoto K. et al., 2008, SoPh, 249, 233
\bibitem[Joos	(2006)]	{joo06}		Joos F. et al., 2006, SPIE, 7016, 48
\bibitem[Kuhn	(1994)]	{kuh94}		Kuhn J.R. et al., 1994, SoPh, 153, 143
\bibitem[Lee	(1998)]	{lee98}		Lee R.L. Jr., 1998, ApOpt, 37, 1465
\bibitem[Lites	(1996)]	{lit96}		Lites B.W., 1996, SoPh, 163, 223
\bibitem[Liu	(1997)]	{liu97}		Liu Y. \& Voss K., 1997, ApOpt, 36, 8753
\bibitem[Mans	(2003)]	{man03}		Manset N. \& Donati J.-F., 2003, SPIE, 4843, 425
\bibitem[Marti	(1999)]	{mar99}		Mart\'{i}nez Pillet V. et al., 1999, ASP Conf. Proc, 183, 264
\bibitem[Moni	(2010)]	{mon10}		Monin D. et al. in prep
\bibitem[Naga	(2007)]	{nag07}		Nagaraju K. et al., 2007, BASI, 35, 307
\bibitem[Pata	(2006)]	{pat06}		Patat F. \& Romaniello M., 2006, PASP, 118, 146
\bibitem[Pomo	(2001)]	{pom01}		Pomozi I. et al., 2001, J. Exp. Biology, 204, 2933
\bibitem[Pove	(2001)]	{pov01}		Povel H.P., 2001, ASP Conf. Proc., 248, 543
\bibitem[Seme	(1993)]	{sem93}		Semel M. et al., 1993, A\&A, 278, 231
\bibitem[Skum	(1997)]	{sku97}		Skumanich A. et al., 1997, ApJS, 110, 357
\bibitem[Snik	(2008)]	{sni08}		Snik F. et al., 2008, SPIE, 7014, 70140O
\bibitem[Sten	(1997)]	{ste97}		Stenflo J.O. et al., 1997, A\&A, 322, 985
\bibitem[Sten	(2007)]	{ste07}		Stenflo J.O., 2007, Mem. S. A. It. 78, 181
\bibitem[Stras	(2003)]	{str03}		Strassmeier K.G. et al., 2003, SPIE, 4843, 180
\bibitem[Stras	(2008)]	{str08}		Strassmeier K.G. et al., 2008, SPIE, 7014, 70140N
\bibitem[Suha	(2004)]	{suh04}		Suhai B. \& Horv\'{a}th G., 2004, JOSA A, 21, 1669
\bibitem[Thor	(2003)]	{tho03}      	Thornton R.J., et al., 2003, Proc. SPIE, 4841, 1115
\bibitem[Tinb	(1997)]	{tin97}		Tinbergen J. \& Rutten R., 1997, ISIS Spectropolarimetry Manual for WHT, www.ing.iac.es/Astronomy/observing/manuals/
\bibitem[Tinb	(2007)]	{tin07}		Tinbergen J., 2007, PASP, 1371, 119
\bibitem[vanH	(2009)]	{van09}		vanHarten G. et al., 2009, PASP, 878, 377
\bibitem[Wolf	(1996)]	{wol96}		Wolff M.J. et al., 1996, AJ, 111, 856
\end{thebibliography}
\end{document}